\documentclass[usenatbib]{mn2e}
\usepackage{epsfig}
\usepackage{rotating}
\usepackage{multirow}
\usepackage{longtable}

\bibpunct[, ]{(}{)}{;}{a}{}{,}

\title[ATLAS Spectroscopic Catalogue \& RLFs]{The Australia Telescope Large Area Survey: Spectroscopic Catalogue and Radio Luminosity Functions}

\author[Mao et al.]{Minnie~Y. Mao$^{1,2,3}$\thanks{e-mail: mymao@utas.edu.au},  Rob Sharp$^{4}$, Ray~P. Norris$^{3}$, Andrew~M. Hopkins$^{2}$,
\newauthor  Nick Seymour$^{3}$, James~E.~J. Lovell$^{1}$, Enno Middelberg$^{5}$,
\newauthor Kate~E. Randall$^{6,3}$,  Elaine~M. Sadler$^{6}$, D.~J. Saikia$^{7,8}$, Stanislav S. Shabala$^{1}$,
\newauthor Peter-Christian Zinn$^{5,3}$\\
$^{1}$School of Mathematics and Physics, University of Tasmania, Private Bag 37, Hobart, 7001, Australia\\
$^{2}$Australian Astronomical Observatory, PO Box 296, Epping, NSW, 1710, Australia\\
$^{3}$CSIRO Australia Telescope National Facility, PO Box 76, Epping, NSW, 1710, Australia\\
$^{4}$Research School of Astronomy and Astrophysics, The Australian National University, Cotter Road, Weston Creek,\\ACT, 2611, Australia\\
$^{5}$Astronomisches Institut, Ruhr-Universit\"at Bochum, Universit\"atsstr. 150, 44801 Bochum, Germany\\
$^{6}$Sydney Institute for Astronomy School of Physics, University of Sydney, NSW 2006, Australia\\
$^{7}$National Centre for Radio Astrophysics, Tata Insitute of Fundamental Research, Pune 411 007, India \\
$^{8}$Cotton College State University, Panbazar, Guwahati 781 001, India
}
\begin{document}

\date{2012}

\pagerange{\pageref{firstpage}--\pageref{lastpage}} \pubyear{200X}

\maketitle

\label{firstpage}

\begin{abstract}
The Australia Telescope Large Area Survey (ATLAS) has surveyed seven square degrees of sky around the Chandra Deep Field South (CDFS) and the European Large Area ISO Survey - South 1 (ELAIS-S1) fields at 1.4\,GHz. ATLAS aims to reach a uniform sensitivity of $10\,\mu$Jy\,beam$^{-1}$ rms over the entire region with data release 1 currently reaching $\sim30\,\mu$Jy\,beam$^{-1}$ rms. Here we present 466 new spectroscopic redshifts for radio sources in ATLAS as part of our optical follow-up program. Of the 466 radio sources with new spectroscopic redshifts, 142 have star-forming optical spectra, 282 show evidence for AGN in their optical spectra, 10 have stellar spectra and 32 have spectra revealing redshifts, but with insufficient features to classify. We compare our spectroscopic classifications with two mid-infrared diagnostics and find them to be in broad agreement. We also construct the radio luminosity function for star-forming galaxies to z~$= 0.5$ and for AGN to z~$= 0.8$. The radio luminosity function for star-forming galaxies appears to be in good agreement with previous studies. The radio luminosity function for AGN appears higher than previous studies of the local AGN radio luminosity function. We explore the possibility of evolution, cosmic variance and classification techniques affecting the AGN radio luminosity function. ATLAS is a pathfinder for the forthcoming EMU survey and the data presented in this paper will be used to guide EMU's survey design and early science papers. 
\\

%The Australia Telescope Large Area Survey (ATLAS) has surveyed seven square degrees of sky around the Chandra Deep Field South (CDFS) and the European Large Area ISO Survey - South 1 (ELAIS-S1) fields at 1.4\,GHz to $\sim30\,\mu$Jy\,beam$^{-1}$ rms.
\end{abstract}

\begin{keywords}
galaxies: active -- galaxies: general -- radio continuum: galaxies -- galaxies: distances and redshifts
\end{keywords}

\section{Introduction}
\label{intro}
Extragalactic radio sources comprise both star-forming (SF) galaxies and active galactic nuclei (AGN) \citep{Condon92}. In order to understand the history of the Universe we must unravel the cosmic evolution of both classes.

The galaxies that host radio sources display a range of different optical spectra. SF galaxy spectra are made up of direct emission from starlight and emission lines from H\textsc{ii} regions, and are quite easily identified at low redshifts. Some radio-loud AGN galaxies display strong high-excitation lines, either broad or narrow depending on obscuration and the orientation \citep{Antonucci93}, while others have weak or no emission features \citep[e.g.][]{Sadler02}. These spectral differences have been attributed to different accretion modes \citep{Hardcastle07, Croton06}. Sources without high-excitation lines are fuelled by the accretion of hot gas (`hot mode') while high-excitation sources require fuelling by cold gas (`cold-mode'). The radiatively inefficient `hot-mode' accretion is driven by the gravitational instability \citep{Pope12}, and results in a geometrically thick, optically thin accretion disk within the hot broad-line region. On the other hand, the radiatively efficient `cold-mode' accretion results in a geometrically thin, optically thick accretion disk within the broad-line region, and a narrow-line region further out. `Cold-mode' accretion may be triggered by a merger event \citep[e.g.][]{Shabala11}, which can also trigger star-formation within the galaxy. Consequently, sources undergoing `cold-mode' accretion may appear as `hybrid' objects, that is, objects whose radio emission results from both star-formation and an AGN component \citep[e.g. PRONGS, ][]{Mao10b}. \citet{Norris11b} find that the extreme ULIRG, F00183-7111, is such an object. 

Early radio surveys such as the 3C and 3CR Radio Surveys \citep{Edge59,Bennett62} had rms levels of a few Janskys and detected predominately powerful radio galaxies and quasars, allowing the cosmic history of these sources to be determined. Subsequent studies found that powerful radio galaxies and quasars evolve very strongly with redshift and the space density of these sources at z $\sim$ 2 is orders of magnitude higher than in the local Universe \citep[e.g.][]{Dunlop90}.

More recent wide-field radio surveys such as NVSS \citep{Condon98}, SUMSS \citep{Mauch03} and FIRST \citep{Becker95} have reached rms levels of a few mJy. At these flux densities the radio source population is dominated by radio galaxies and quasars powered by AGN \citep[e.g.][]{Sadler02}. The most powerful of these can be seen out to the edge of the visible Universe.

Source counts show a well-characterised upturn below 1\,mJy\,beam$^{-1}$, which is above that predicted from the extrapolation of source counts for AGN with high-flux densities. Some studies attribute this to strong evolution of the SF galaxy luminosity function \citep[e.g.][]{Hopkins98} while others find a significant contribution from lower-luminosity AGN \citep[e.g.][]{Huynh07}. \citet{Sadler07} found that low-luminosity AGN undergo significant cosmic evolution out to z = 0.7, consistent with studies of the optical luminosity function for AGN \citep[e.g.][]{Croom04}.

The deepest radio surveys now reach rms levels below 10\,$\mu$Jy\,beam$^{-1}$ \citep[e.g.][]{Owen08}, but cover very small areas on the sky. The distribution of AGN and SF galaxies at low flux densities is not well known, with some studies finding the proportion of AGN declining with decreasing flux density and emission due to star-formation (SF) dominating \citep[e.g.][]{Mauch07, Seymour08}, while other studies find the distribution is closer to a 50/50 split between SF galaxies and AGN  down to $\sim$50\,$\mu$Jy\,beam$^{-1}$ \citep[e.g.][]{Smolcic08, Padovani09}. The lack of consensus of the distribution of AGN and SF at low flux densities is due to the small areas of sky surveyed at these sensitivities. \citet{Norris11a} provide a brief comparison of the various studies and, in their figure 4, show that the fraction of SF galaxies at $<$100\,$\mu$Jy\,beam$^{-1}$ could range from 30 - 80 per cent.

To date, understanding the evolution of the faint radio source population has been limited by the availability of sufficiently deep wide-area radio surveys. With the aim of imaging approximately seven square degrees of sky over two fields to 10\,$\mu$Jy\,beam$^{-1}$ at 1.4\,GHz, the Australia Telescope Large Area Survey \citep[ATLAS,][]{Norris06,Middelberg08} is the widest deep field radio survey yet attempted. ATLAS observes two separate fields so as to minimize cosmic variance \citep{Moster11}: Chandra Deep Field South (CDFS) and European Large Area ISO Survey-South~1 (ELAIS-S1). The two fields were chosen to coincide with the \textit{Spitzer} Wide-Area Infrared Extragalactic (SWIRE) Survey program \citep{Lonsdale03}, so optical and mid-infrared identifications exist for most of the radio objects. CDFS also encompasses the Southern Great Observatories Origins Deep Survey (GOODS) field \citep{Giavalisco04}. 

The first data release (Data Release 1, or DR1) from ATLAS consists of the preliminary data published by \citet{Norris06} and \citet{Middelberg08} and reaches an rms sensitivity of $\sim$30$\,\mu$Jy\,beam$^{-1}$. The flux densities of DR1 are accurate to $\sim$10 per cent. The final ATLAS data release (DR3: Banfield et al., in preparation) will include additional data taken with the new CABB upgrade \citep{Wilson11} to the ATCA, and is expected to reach an rms sensitivity of $\sim$10$\,\mu$Jy\,beam$^{-1}$. For this paper we use radio data only from DR1.

Spectroscopic data provide accurate redshift information that can be used to determine absolute magnitudes, luminosities etc. This paper presents spectroscopic redshifts and spectroscopic classifications of radio sources in the ATLAS fields obtained using the AAOmega multi-fibre spectrograph at the Anglo-Australian Telescope (AAT) of sources in the ATLAS fields. Section 2 describes the observations and data analysis while Section 3 provides spectroscopic classifications and a description of the redshift catalogue. Section 4 discusses how the spectroscopic classifications compare to mid-infrared diagnostics and presents the radio luminosity function for ELAIS. Section 5 summarises our conclusions.

This paper uses H$_0$ = 70 km s$^{-1}$ Mpc$^{-1}$, $\Omega$$_M$ = 0.3
and $\Omega$$_\Lambda$ = 0.7 and the web-based calculator of
\citet{Wright06} to estimate the physical parameters. Vega magnitudes are used throughout.
   
\section{Observations and Data}\label{data}

\subsection{Existing multi-wavelength data in ATLAS}
Radio observations of ATLAS were carried out with the Australia Telescope Compact Array (ATCA, Project ID C1241). The ATCA was used in a variety of configurations (6A, 6B, 6C, 6D, 1.5D, 750A, 750B, 750D, EW367) to maximise $uv$ coverage. ATLAS observations were performed in mosaic mode with 40 pointings. The radio data used in this paper are from DR1 \citep{Norris06, Middelberg08}. CDFS was observed for 173 hours in total, or 8.2 hours per pointing with a resolution of $11'' \times 5''$, and ELAIS was observed for 231 hours in total, or 10.5 hours per pointing with a resolution of $10.3'' \times 7.2''$. There are 2018 radio sources in ATLAS within a total area of 7.65 square degrees. Smaller subsets of the ATLAS-CDFS field have been observed using the VLA by \citet{Kellermann08} and \citet{Miller08} at 1.4\,GHz, but these data are not used here. Cross-matching of the ATLAS radio sources to optical and infrared counterparts is discussed by \citet{Norris06} and \citet{Middelberg08}. 

Infrared observations of ATLAS were obtained by SWIRE \citep{Lonsdale03}, the largest \emph{Spitzer} legacy program. SWIRE covers seven different fields including CDFS and ELAIS, totalling 60 - 65 square degrees in area at all seven Spitzer bands. SWIRE observes at 3.6, 4.5, 5.8, 8 and 24\,$\mu$m down to 5--10 sigma detection limits (band-dependent) of 5, 9, 43, 40 and 193\,$\mu$Jy respectively\footnote{http://swire.ipac.caltech.edu/swire/swire.html}. The SWIRE data used in this paper are all from the third SWIRE data release.

\subsection{Anglo-Australian Telescope Observations and Data Reduction}

\subsubsection{Source Selection}
The aim of the optical observations was to obtain spectra for all radio sources. A small fraction ($< 10\%$) have spectroscopic redshift information in the literature, predominately in CDFS. Consequently, these were assigned a low observation priority. Furthermore, radio sources with very faint or no optical counterparts were also assigned a low observation priority.

%The position of counterparts for optical spectroscopy were identified with the SWIRE mid-infrared positions to provide sub-arcsecond astrometry \citep{Norris06,Middelberg08}. Optical photometry at these positions was derived from SWIRE\footnote{http://swire.ipac.caltech.edu/swire/swire.html} \citep{Lonsdale03} to identify our sample (both `bright' and \textbf{`intermediate'} in Table \ref{obs}) consisting of sources with $16<R<22$, which were suitable for normal AAOmega observations. The optical photometric data in SWIRE is not uniform, with areas where no data are available. Within these areas we used photometric data from SuperCOSMOS \citep{Hambly01}.  

The position of counterparts for optical spectroscopy were identified with the SWIRE mid-infrared positions to provide sub-arcsecond astrometry \citep{Norris06,Middelberg08}. Optical photometry at these positions was derived from SWIRE \citep{Lonsdale03} to identify our sample in Table \ref{obs}). The optical photometric data in SWIRE is not uniform, with areas where no data are available. Within these areas we used photometric data from SuperCOSMOS \citep{Hambly01}.  

The target sources were separated into `bright', `intermediate' and `faint' sub-samples. The bright and intermediate sub-samples, with $R<20$ and $20<R<22$ respectively, were suitable for normal AAOmega observations. The `bright' sample yielded the majority of the successfully determined redshifts presented in this paper. In addition,  the `faint' sample of sources  with $R>22$ was selected for nod+shuffle observing, but these were unable to be completed because of poor weather. Of the 1120 sources observed in total, 692 were in the `bright' sample,  which yielded high quality redshifts for 466 sources. 

\subsubsection{Observation}

\begin{table*}
\begin{center}
\caption{Summary of AAT observations. N. obs gives the number of fibres placed on science sources. The seeing information is from the AAT observing log. All observations were performed in multi-object mode with the exception of `CDFS A ns' and `CDFS B ns', which were performed in nod+shuffle mode. An unspecified Seeing value indicates an unavailable measurement. }\label{obs}
\scriptsize{
\begin{tabular}{lccccc}
\\
\hline
Observation & N. obs. & Exposure (s) & Obs. date & Dichroic & Seeing (arcseconds)\\
\hline
CDFS A bright & 329 & 3 $\times$ 1200& 20071202 & 5700\AA\ & 2.4 \\
CDFS A bright & 343 & 4 $\times$ 1200 & 20071204 & 5700\AA\ & \ldots \\
CDFS B bright & 329 & 4 $\times$ 1200 & 20071204 & 5700\AA\ & \ldots\\
\\
CDFS A intermediate & 343 & 4 $\times$ 1800 & 20071207 & 5700\AA\ & 1.9-2.5 \\
CDFS B intermediate & 329 & 2 $\times$ 1200 & 20071205 & 5700\AA\ & 1.8-2.3\\
\\
CDFS A ns & 162 & 1 $\times$ 1500 & 20071206 & 5700\AA\ & 1.4 \\
CDFS A ns & 162 & 13 $\times$ 2400 & 20071206,07,08 & 5700\AA\ & 1.2-3 \\
CDFS A ns & 187 & 5 $\times$ 2400 & 20071208 & 5700\AA\ & 1.2 - 1.7\\
CDFS B ns & 187 & 1 $\times$ 1500 & 20071205 & 5700\AA\ & 2.3\\
\\
CDFS A 2010 & 329 & 6 $\times$ 2400 & 20101206 & 5700\AA\ & \ldots \\
CDFS B 2010 & 329 & 2100+1500 & 20101207 & 5700\AA\ & \ldots\\

\hline
ELAIS bright & 343 & 7 $\times$ 1200 & 20071205 & 5700\AA\ & 1.8\\
ELAIS bright & 343 & 4 $\times$ 1800 & 20081221,22 & 6700\AA\ & 1.6-1.7 \\
ELAIS bright & 329 & 4 $\times$ 1800  & 20081224,25,26 & 6700\AA\ & 1.6-2.4 \\
\\
ELAIS 2010 & 343 & 1270+1770 & 20101208 & 5700\AA\ & 1.4 - 1.6\\
\hline
\end{tabular}
}
\end{center}
\end{table*}

We obtained AAOmega observations of the ATLAS radio sources from December 1 to December 8, 2007 and December 20 to December 27, 2008, all of which were dark nights. AAOmega is the multi-object spectrograph on the Anglo-Australian Telescope (AAT). We also obtained some observations for ATLAS sources from December 1 to December 8, 2010 as part of a complementary project. The AAOmega spectrograph was used in multi-object mode \citep{Saunders04, Sharp06} and nod+shuffle mode \citep{Glazebrook01}. While the sensitive principle component analysis (PCA) based approach for deep observations with AAOmega \citep{Sharp10} would likely have resulted in enhanced sensitivity, the technique was still under development at the time of our observations. We used the dual beam system with the 580V and 385R Volume Phase Holographic gratings. The 2007 and 2010 observations used the 5700\AA\ dichroic beam splitter and covered the spectral range between 3700\AA\ and 8500\AA\ at central resolutions in each arm of R$\sim$1300 per 3.4 pixel spectral resolution element. The 2008 observations used the 6700\AA\ dichroic beam splitter and covered the spectral range between 4700\AA\ and 9500\AA, which allows detection of H$\alpha$ to z~$\sim$~0.4.

Each of the three observing sessions over three years was compromised by poor weather, with only $16.9\%$ of scheduled time able to be used, severely restricting the sensitivity of the observations. The seeing ranged from 1.2$''$ to 3.0$''$. The observations are summarised in Table \ref{obs}. 

The sensitivity limit (SNR$= \sim$3 per-pixel) with AAOmega for our estimated exposure times was $R\sim22$. The surface density of ATLAS sources at this sensitivity limit is $\sim$100/deg$^{2}$, with this value decreasing markedly for brighter sensitivity limits. Optimising the placement of the fibres still resulted in `spare fibres' that could not be placed on ATLAS radio sources. The remaining fibres were allocated to three ATLAS related projects: the detection of a cluster associated with a wide-angle-tail galaxy, a sample of luminous red galaxies and a large sample of 24\,$\mu$m excess sources\footnote{Sources that have a high 24\,$\mu$m to radio flux density ratio, log(S$_{24\mu\rm{m}}$/S$_{20\rm{cm}}$)~$>$~1.25,  are defined as 24\,$\mu$m excess sources. This criterion was determined empirically.} in the ATLAS fields. The redshifts obtained for the cluster work have been published in \citet{Mao10a} and the 24\,$\mu$m data are presented in Appendix \ref{24umsources} and will be discussed in further detail in Norris et al. (in preparation), but are not discussed further in the context of this work.

\subsubsection{Data Reduction}
The spectroscopic data were processed using the 2dfDR software provided by the AAO. Source redshifts were determined using the \textsc{runz} redshift-fitting package using template cross-correlation or emission line fitting with a redshift quality flag out of 5 assigned via visual inspection of the fit\footnote{\textsc{runz} was developed for the 2dFGRS survey \citep{Colless01} by Will Sutherland, and is based on the \citet{Tonry79} cross-correlation technique. The code was further developed by Will Saunders, Russell Cannon, Scott Croom and others for a number of additional surveys undertaken with the AAOmega instrument at the AAT}. Spectra with a quality flag of 2 or below were discarded. 

621 ATLAS radio sources were observed in ELAIS, with spectra of quality $> 2$ obtained for 306 sources. 499 ATLAS radio sources were observed in CDFS, with spectra of quality $> 2$ obtained for 160 sources. A total of 1120 ATLAS radio sources were observed, with spectra obtained for 466 sources yielding reliable redshifts from the AAOmega observations. 

\begin{figure}
\begin{center}
\includegraphics[angle=0, scale=0.8]{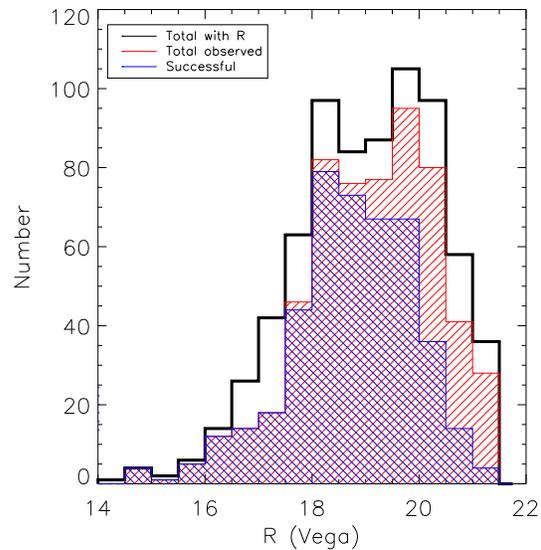}
\end{center}
\caption{Histogram of $R$-band magnitudes for the target sources. The black histogram represents all ATLAS radio sources that have $R$-band magnitudes from SuperCOSMOS \citep{Hambly01}, the red histogram represents the sources that were observed with AAOmega and the blue histogram represents all sources that had quality $> 2$ spectra. }\label{optcompall}
\end{figure}

Our observations achieve $\sim70\%$ completeness at $R\sim20$ (Figure \ref{optcompall}). This corresponds to the ability to detect typical SF galaxies to z~$\sim 0.4$ and typical early-type galaxies to z~$\sim 0.5$.% The bright ($\sim$R~$<18$) sources that were not observed either already had redshift information in the literature or were no

%Despite attempting to observe 1120 ATLAS radio sources, many of these were unsuccessful as they had faint R-band magnitudes, or no magnitude information available, and were observed as part of the nod+shuffle observations, which were unable to be completed due to poor weather. It may be fairer to consider that of the 1120 sources observed, 692 were observed as part of the `bright' observations, and the majority of the successfully determined redshifts were from the `bright' subset. That is to say, 692 ATLAS radio sources brighter than R~$\sim$22 were observed with AAOmega, with quality spectra obtained for 466 yielding reliable redshifts. 

%Data obtained with conventional mode was reduced using the standard 2dfDR routines DRCONTROL. The data from the blue and red arms was then SPLICED together. %FIND OUT MEAN S:N

%The spectra was analyzed using the redshift extraction package RUNZ. Redshifts were measured using this program and manually assessed for reliability. The spectra were also classified at this stage.

%The script to generate the tables is here:
%`~/phd/atlas\_spectroscopy/zmatch\_lit.pro'
%HERE NOW
%`/luminosity/lum\_SFvAGN/lumfn\_SFvAGN.pro'

%~/phd/atlas_spectroscopy/zmatch_lit.pro

%All ATLAS sources in CDFS. The Rmags need to be checked as I've currently only included Rmags that were available on the source export page - but I didnt carry across whether it was not observed OR if it was observed unsuccessfully. The redshifts should all be right. There are a total of 265 sources with redshift information, of which 160 are from our own observations. 

%\subsection{CDFS}

\section{Results}
%1478 sources observed in CDFS. 826 yielded successful redshifts. 499 ATLAS sources were observed, 160 of which had quality redshifts. 633 non-ATLAS (24$\,\mu$m) sources were observed of which 469 had quality redshifts. 346 non-ATLAS (Rob's AGN, LRGs etc) were observed of which 197 had quality redshifts.

%722 sources observed in ELAIS. 337 yielded successful redshifts. 621 ATLAS sources were observed, 306 of which had quality redshifts. 63 non-ATLAS sources were observed of which 11 had quality redshifts. These sources were putative cluster members around S1189 for our WATs in ATLAS project. 38 non-ATLAS sources (Rob's AGN and LRGs) were observed of which 20 had quality redshifts. 

%We observed a total of 1120 ATLAS sources and obtained quality spectra for 466 of the ATLAS sources from our AAT observations. We also observed a total of 1080 non-ATLAS sources of which 697 yielded quality spectraWHAT DID ROB SAY ABOUT THIS - CHECK HIS NOTES. The 24$\,\mu$m excess sources are currently being studied in the context of sources that deviate off the radio-FIR correlation in Norris et al. (in prep). The redshifts for the non-ATLAS sources are presented in Appendix X. For the remainder of this paper we will only be considering the radio ATLAS sources. 

\subsection{Spectroscopic Classifications}\label{sc}
In the course of measuring the redshift, each spectrum was inspected visually to determine whether the dominant physical process responsible for the radio emission was star-formation (SF) or an AGN. This method is similar to that used by previous studies \citep[e.g][]{Sadler02}. \citet{Sadler99} reported that visual classifications can be used with confidence to analyse spectra from 2dF, the precursor to AAOmega. 

\begin{figure*}
\begin{center}
\begin{tabular}{cc}
\includegraphics[angle=0, scale=0.5]{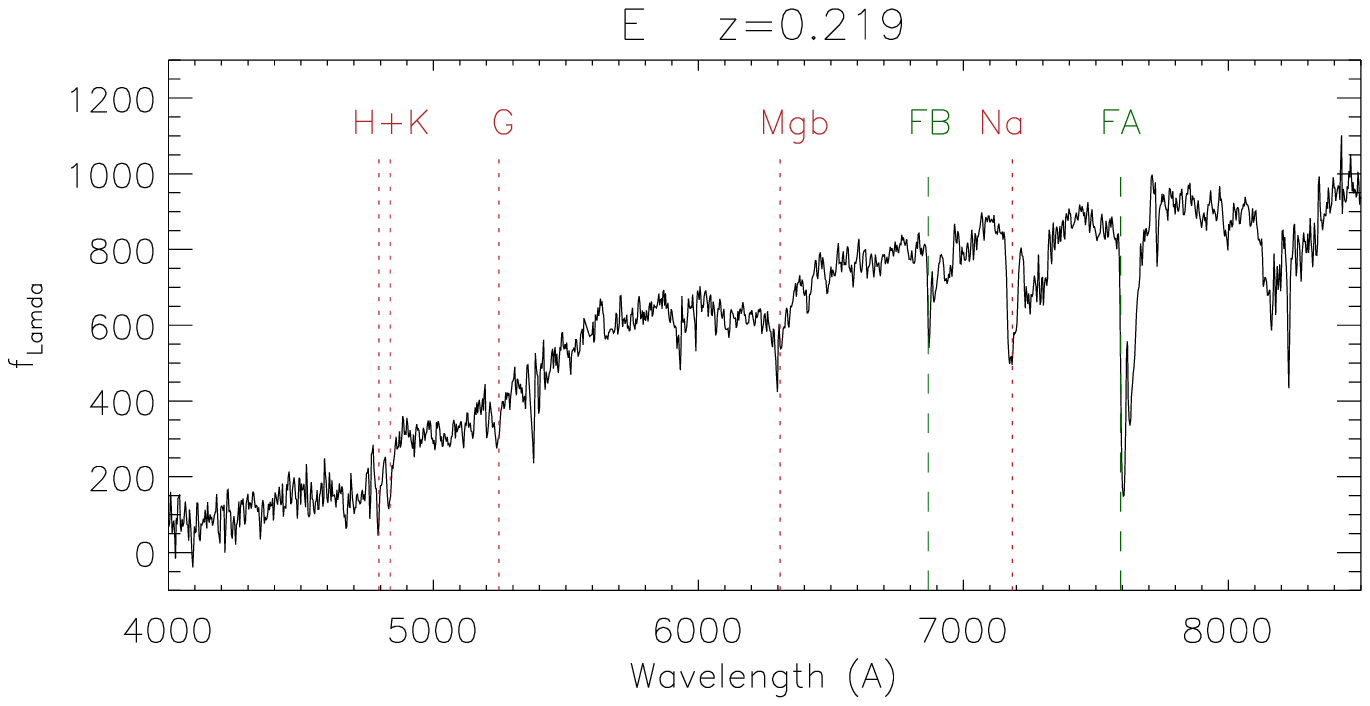}& \includegraphics[angle=0, scale=0.5]{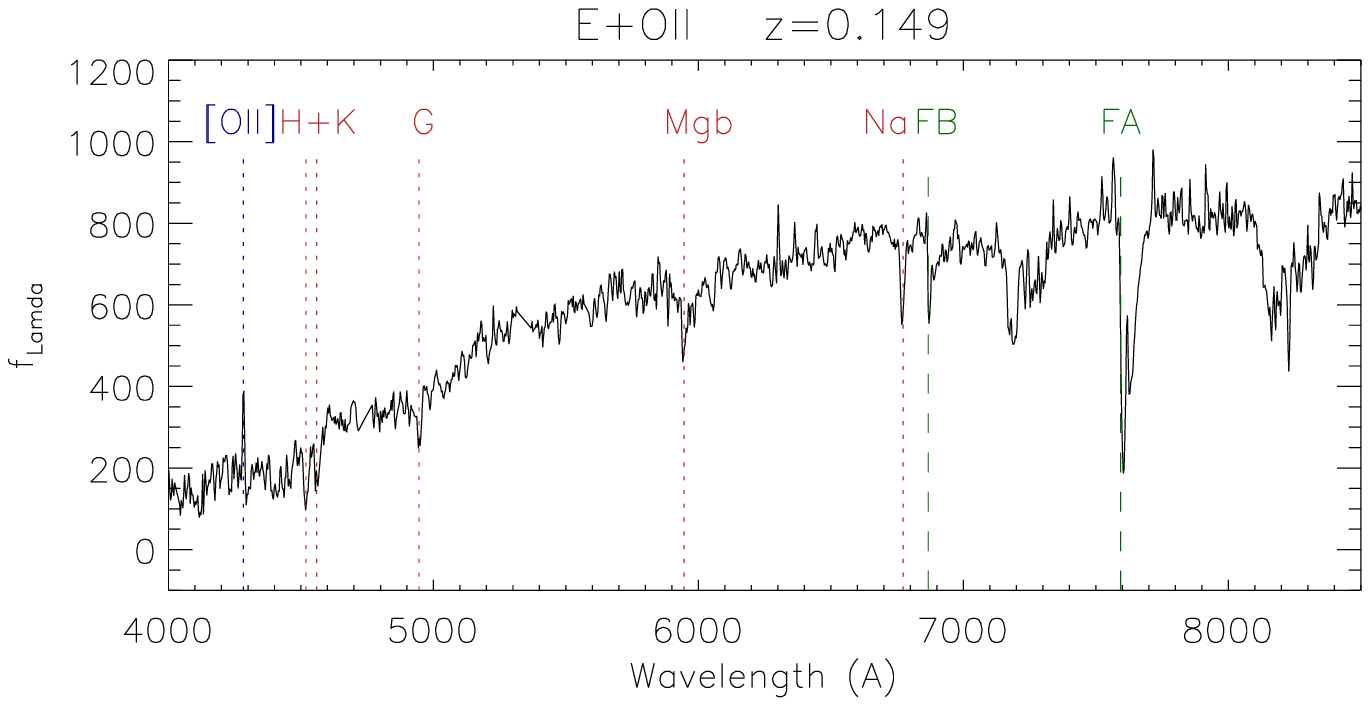} \\
\includegraphics[angle=0, scale=0.5]{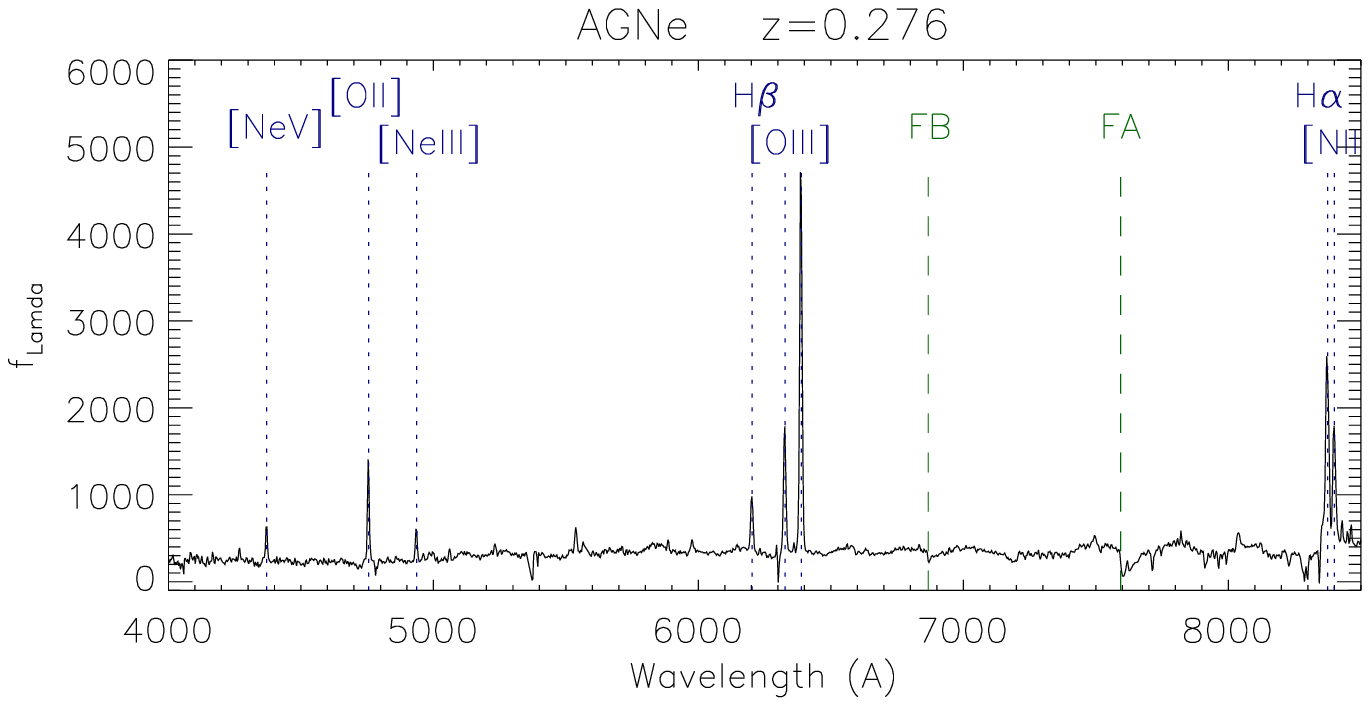} & \includegraphics[angle=0, scale=0.5]{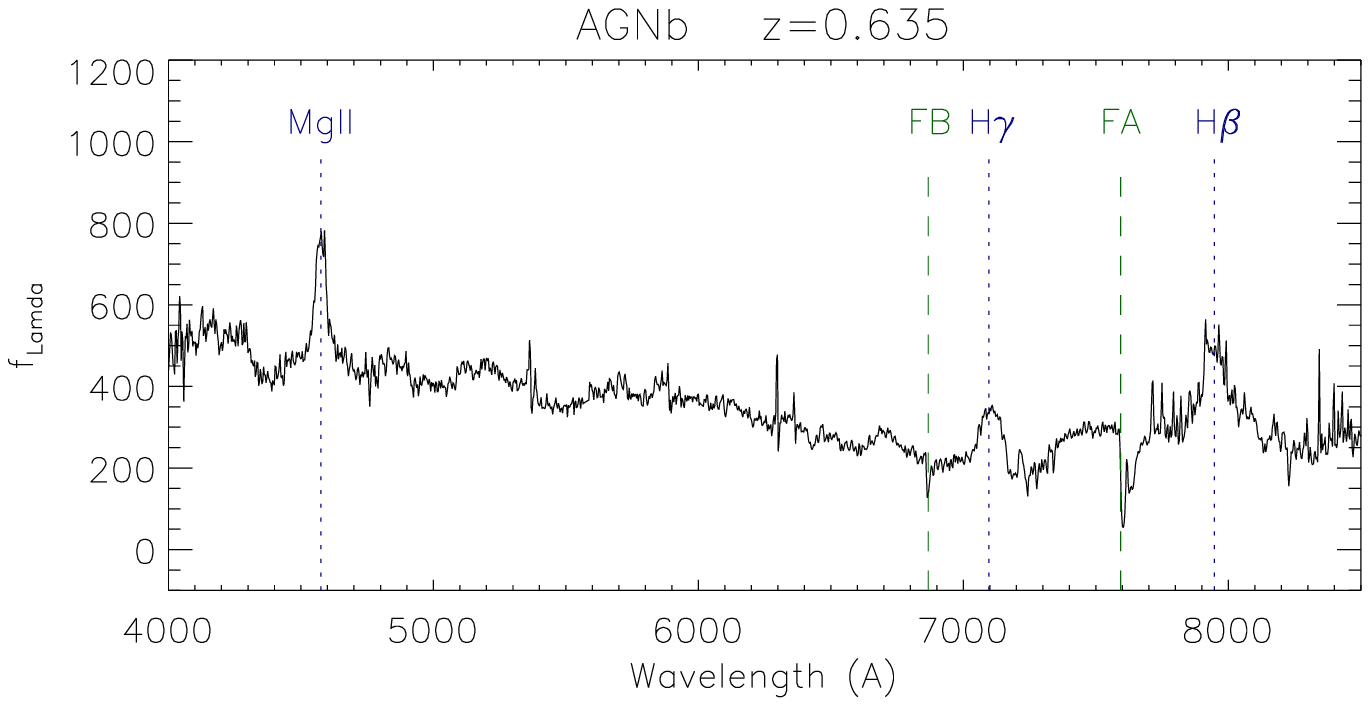}\\
\includegraphics[angle=0, scale=0.5]{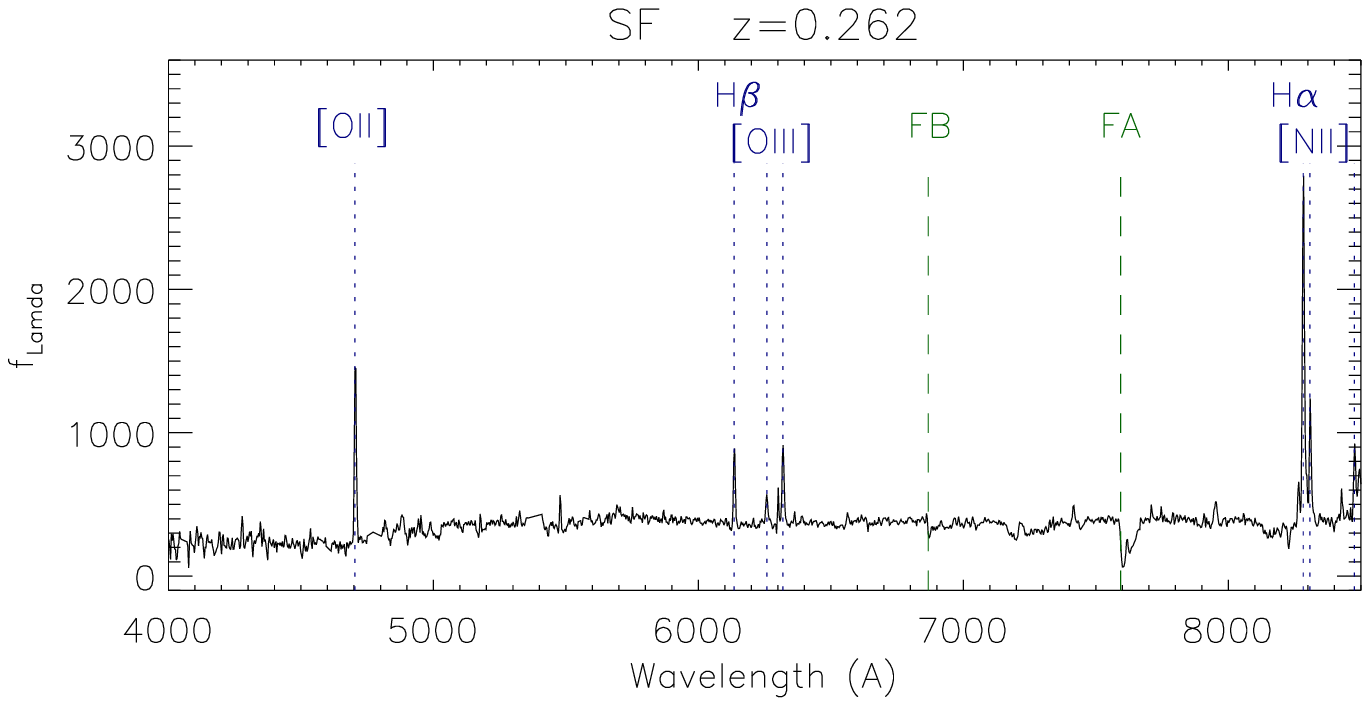}\\
\end{tabular}
\end{center}
\caption{Examples of each spectroscopic classification. From top left the spectra are `E', `E+OII', `AGNe', `AGNb' and `SF'. The red dotted lines indicate prominent absorption features and the blue dotted lines indicate prominent emission features. The green dashed lines indicate the Fraunhofer A+B atmospheric absorption bands from O$_2$.}\label{spec}
\end{figure*}

SF galaxy spectra are typically dominated by strong, narrow emission lines including the Balmer series. We classified these spectra as `SF'. AGN, on the other hand, can have pure absorption-line spectra (`E'), absorption-line spectra with some low-ionisation emission lines such as [OII] (3726\,\AA, 3729\,\AA) (`E+OII'), emission-line spectra whose line ratios are indicative of AGN activity (`AGNa') and broad-line spectra (`AGNb'). Figure \ref{spec} provides examples of each type of spectrum. Ten sources had stellar spectra and, upon visual inspection of the field, we attribute this to chance alignments and discard these data from further analysis in this work.

There are 142 SF, 282 AGN (110 E, 60 E+OII, 79 AGNe, 32 AGNb), 10 stars and 32 `unknown' spectra that we are unable to spectroscopically classify, despite sufficient features for a redshift to be determined. The spectra that are classified as `unknown' are typically at z~$>$~0.4, so H$\alpha$ is redshifted out of the spectrum. Furthermore, a ratio of [OIII]/H$\beta$~$\sim$~1 could either indicate SF or AGN, and without further information spectroscopic classifications cannot be made. We acknowledge the presence of hybrid systems in our sample, but do not make any attempt to classify such sources in this paper and defer this to a future paper. 

\subsection{Redshift and Spectroscopic Classification Catalogue}

The catalogue of redshifts and spectroscopic classifications is available as supplementary material in the online version of this paper. Table \ref{zcat} provides the first ten lines of the catalogue. 

\begin{table*}
\begin{center}
\caption{First 10 lines of the catalogue of new spectrosopic redshifts and spectroscopic classifications of the 466 radio-selected galaxies in ATLAS. Column 1 gives the `SID' from \citet{Norris06,Middelberg08} and Columns 2 and 3 provide the position of the optical counterpart. The radio flux densities in Column 6 are from \citet{Norris06,Middelberg08} and are uncorrected for effects such as bandwidth smearing. Columns 5 and 6 give the SuperCOSMOS $R$ and $B$-band magnitudes \citep{Hambly01}. Columns 7 and 8 present the spectroscopic redshifts and spectroscopic classifications determined in this paper and Columns 9 and 10 present the absolute $R$-band magnitude and 1.4\,GHz radio luminosity. }\label{zcat}
\begin{tabular}{cllrrrcrrr}
\\
\hline

SID  & RA & Dec & S20 & $R$ & $B$ & z & Spec. class. & M$_{R}$ & log L$_{1.4\,GHz}$\\
& (J2000) & (J2000) & (mJy) & (Vega) & (Vega) & & & & (W~Hz$^{-1}$)\\
\hline
S007& 03:26:15.42&-28:46:30.80&        0.71&       20.19&       22.52&      0.7108&E+OII&      -23.68&       24.15\\
S009& 03:26:16.32&-28:00:14.72&        1.66&       20.46&      \ldots&      0.5295&E+OII&      \ldots&       24.22\\
S012& 03:26:22.06&-27:43:24.53&       27.81&       19.12&       19.37&      1.3871&AGNb&      -25.01&       26.42\\
S014& 03:26:26.90&-27:56:11.65&        4.14&       17.98&       18.92&      0.8106&AGNb&      -25.46&       25.05\\
S015& 03:26:29.13&-28:06:50.80&        0.31&       16.66&       17.21&      0.0579&SF&      -20.44&       21.40\\
S021& 03:26:30.65&-28:36:58.03&        1.86&       18.98&       21.54&      0.4731&E&      -23.89&       24.15\\
S031& 03:26:39.12&-28:08: 1.57&       42.29&       16.89&       18.43&      0.2184&E&      -23.55&       24.75\\
S037& 03:26:43.38&-28:13:28.06&        0.50&       17.93&       19.79&      0.2956&SF&      -23.39&       23.12\\
S038& 03:26:43.34&-28:22:11.43&        8.28&       19.37&       21.69&      0.3220&AGNe&      -22.32&       24.42\\
S042& 03:26:48.47&-27:49:38.06&        5.00&      \ldots&      \ldots&      2.9869&AGNb&      \ldots&       26.43\\

\hline
\end{tabular}
\end{center}
\end{table*}

%The redshift histograms are generated here:
%`~/phd/atlas\_spectroscopy/zmatch\_atlas.pro'

%~/phd/atlas_spectroscopy/zmatch_atlas.pro

%IDL> print, median(zfin[haszcdfs])                
%     0.331800
%IDL> print, median(zfin[haszelais])
%     0.315090

\begin{figure}
\begin{center}
\includegraphics[angle=0, scale=0.8]{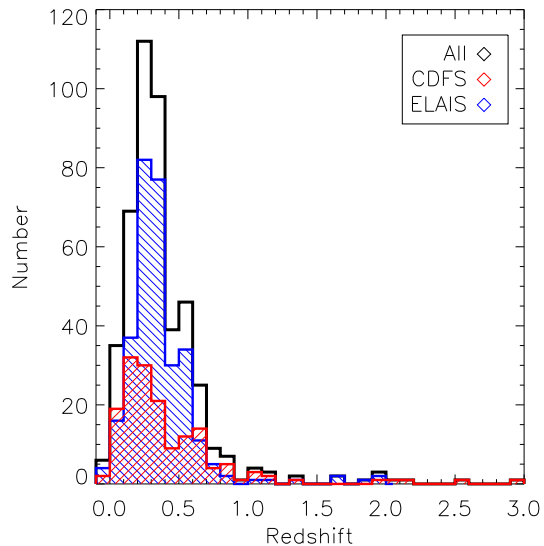}
\end{center}
\caption{Histogram of spectroscopic redshifts in ATLAS. The red line shows data for CDFS and the blue line shows data for ELAIS. The black line is the total. The bins are 0.1\,z wide.}\label{zhist}
\end{figure}

Figure \ref{zhist} shows the redshift distribution of the 466 sources in ATLAS for which we obtained redshifts from AAOmega. The median redshift for the entire sample is 0.316 while the median redshift for CDFS is 0.332 and the median redshift for ELAIS is 0.315. The redshift peak in the CDFS redshift histogram at 0.6 $<$ z $<$ 0.7 corresponds to a known cosmic sheet \citep[e.g.][]{Norris06}. There is also a redshift peak in the ELAIS redshift histogram at $0.2 <$~z~$<0.4$. \citet{Mao10a} discovered an overdensity in ELAIS at z~$\sim 0.22$, associated with a large wide-angle tail galaxy (WAT). They also found an additional three WATs in ELAIS at $0.3<$~z~$<0.4$, each of which is likely to be associated with a cluster of galaxies. %The redshift peak in the ELAIS redshift histogram at 0.2 $<$ z $<$ 0.3 corresponds to an overdensity associated with a large wide-angle tail galaxy \citep{Mao10a}. 

The most distant radio source in our observed sample is SWIRE3 J032648.47-274938.0 at a redshift of z = 2.99, and has been spectroscopically confirmed as a broad-line quasar with a radio power of 2.8~$\times$~10$^{26}$~W~Hz$^{-1}$. 

We calculate the radio luminosity for each source with spectroscopic redshift information using 

\begin{equation}
{\rm L}_\nu=\frac{4\pi{\rm D}_L^2{\rm S}_{\nu obs}}{(1+z)^{1+\alpha}}
\end{equation}

where L$_\nu$ is the luminosity in W~Hz$^{-1}$ at the frequency $\nu$,
D$_{L}$ is the luminosity distance in metres, 
S$_{\nu obs}$ is the observed flux density at the observing frequency, and
$\alpha$ is the spectral index where spectral index is defined as S $\propto$ $\nu^{\alpha}$. We assume $\alpha$~=~-0.75, which is a typical mean value for the synchrotron emission observed from faint radio sources \citep{Ibar10}, both star-forming and AGN. 

\begin{figure}
\begin{center}
\includegraphics[angle=0, scale=0.8]{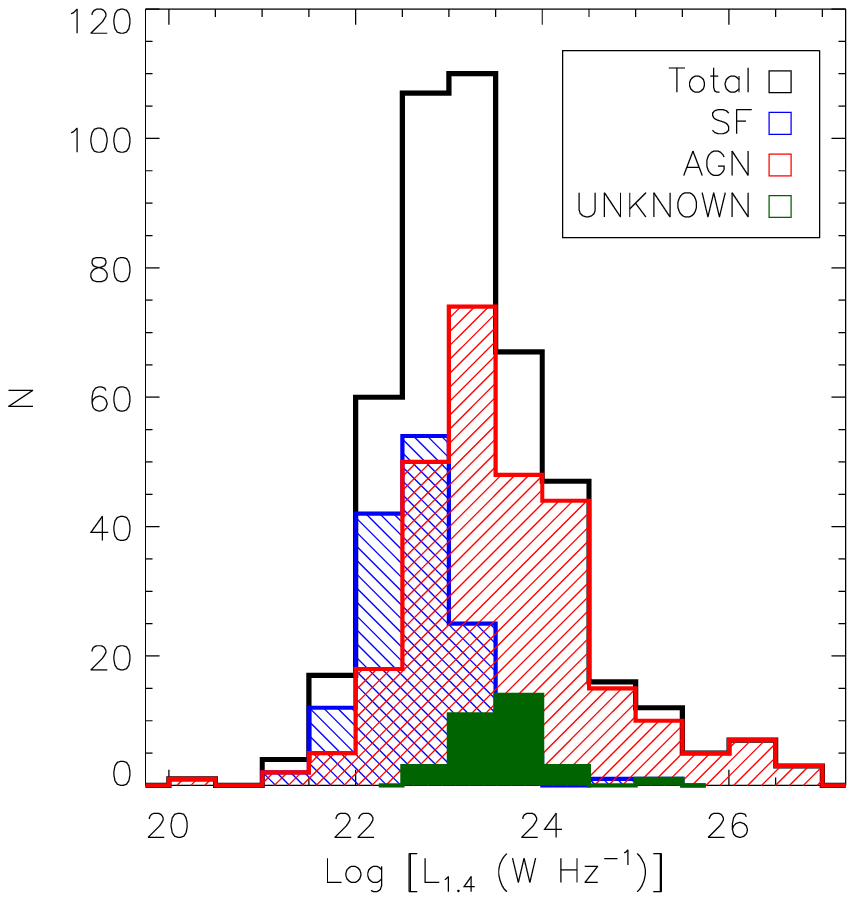}
\end{center}
\caption{Radio luminosity histogram for our sample. Sources spectroscopically classified as SF are shown in blue, while those classified as AGN are shown in red. Those without spectroscopic classifications are shown in green. }\label{lumhist}
\end{figure}

Figure \ref{lumhist} presents the luminosity histogram for sources with redshift information. The median L$_{1.4\, \rm GHz}$ for these sources is $\sim 10^{23.1}$\,W\,Hz$^{-1}$ with the SF source median L$_{1.4\, \rm GHz} \sim  10^{22.6}$\,W\,Hz$^{-1}$ and the AGN median L$_{1.4\,\rm GHz} \sim 10^{23.4}$\,W\,Hz$^{-1}$. The large overlap between the radio luminosity ranges for SF and AGN sources makes it impossible to determine whether sources with the `unknown' spectroscopic classification are SF or AGN from this information alone.

\section{Discussion}

\subsection{Mid-IR colours}

We compare our spectroscopic classifications with those determined from mid-infrared colour-colour diagrams similar to those of \citet{Lacy04} using the SWIRE data from \emph{Spitzer}. Not all of the 466 sources with spectroscopic classifications have infrared data at the SWIRE wavelengths and, as such, comparisons may only be made where data is available at all wavelengths. 

\begin{figure}
\begin{center}
\includegraphics[angle=0, scale=0.8]{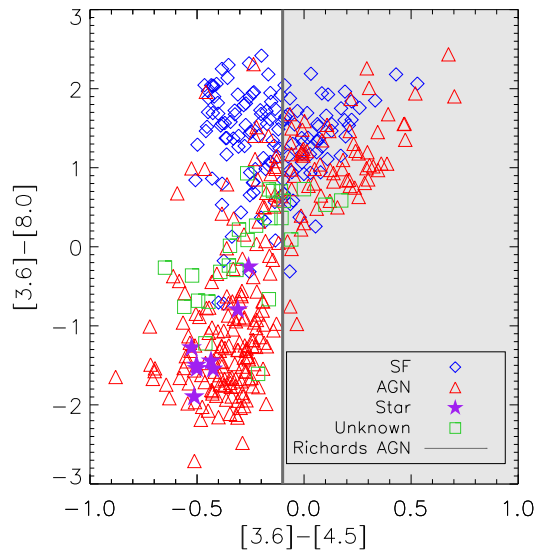}
\end{center}
\caption{MIR $[3.6\,\mu \rm m]-[4.5\,\mu$m] vs. $[3.6\,\mu \rm m]-[8\,\mu$m] colour-colour plot. The data in blue show the sources classified spectroscopically as SF and those in red show the sources classified as AGN. Stars are shown in purple and the sources we were unable to classify spectroscopically are shown in green. The grey shaded region shows the location of the \citet{Richards06} selection for Type 1 quasars.}\label{ir_cols1}
\end{figure}

Figure \ref{ir_cols1} shows the $[3.6\,\mu \rm m]-[4.5\,\mu \rm m]$ against $[3.6\,\mu \rm m]-[8\,\mu$m] plot, where $[1]-[2]=-2.5$log(S$_1$/S$_2$), that is, the SWIRE colours are given in AB magnitude units to allow direct comparison with \citet{Richards06}. The spectroscopic classifications are plotted in different colours. The demarcation between AGN and SF galaxies is as expected \citep[e.g][]{Sajina05}, which provides an independent validation of our visual spectroscopic classification method. 

%distribution of the AGN and SF galaxies are REWORD pretty much as expected which provides an independent validation of our REWORD eyeballing method. %Figure \ref{ir_cols2} shows a similar plot using an extra IR band...

%The sources for which we spectroscopically classified as `unknown' (green points) are located in the area of the plot where there exist both SF and AGN sources (`the valley of despair'). Consequently, these sources with `unknown' spectroscopic classification cannot be classified as SF or AGN based on their mid-infrared colours either. 

\citet{Richards06} suggest that Type 1 quasars may be selected by taking $[3.6]-[4.5] > -0.1$. We apply this selection to our data and find 127 sources would be classified as Type 1 AGN (Figure \ref{ir_cols1}). Of these 127 sources, we had spectroscopically classfied 55 as SF, 67 as AGN (27 AGNb, 36 AGNe, 4 E or E+OII) and 5 were `unknown'. While this method appears to select against E type sources more robustly than the \citet{Lacy04} method below, the contamination by SF sources is considerable.  

\begin{figure}
\begin{center}
\includegraphics[angle=0, scale=0.8]{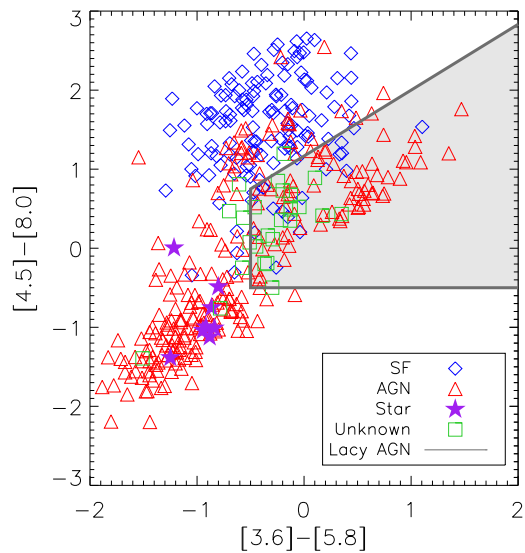}
\end{center}
\caption{MIR $[3.6\,\mu \rm m]-[5.8\,\mu$m] vs. $[4.5\,\mu \rm m]-[8\,\mu$m] colour-colour plot. The data in blue show the sources classified spectroscopically as SF and those in red show the sources classified as AGN. Stars are shown in purple and the sources we were unable to classify spectroscopically are shown in green. The grey shaded region shows the location of the \citet{Lacy04} selection for AGN.}\label{ir_cols2}
\end{figure}

\citet{Lacy04} imposed a MIR colour selection to select AGN in their
sample of SDSS sources in the First Look Survey. Applying this selection criteria to the ATLAS data we classify 100 sources as AGN (Figure
\ref{ir_cols2}). Of the 100 AGN that satisfied the `Lacy' AGN
criteria, we had spectroscopically classified 18 as SF, 65 as AGN (26 AGNb,
27 AGNe, 12 E or E+OII) and 17 were `unknown'. Our spectroscopic
classifications show that there is a significant region of overlap
between SF and AGN sources on this MIR colour-colour plot. Indeed, the
majority of the `unknown' spectroscopic classifications fall in this
region of overlap. A more stringent `diagonal boundary' on the AGN
selection may be better for AGN selection, but this would also lose
many AGN in the overlap region. We note that \citet{Lacy07} shifted
the vertical boundary of the `Lacy' wedge slightly but this does not
affect our result significantly. %Preliminary analysis indicates that
%this method is not more likely to pick up broad-line AGN (Type 1) over
%emission-line AGN and this will be explored in more detail in a future paper.

% This criterion is best suited for the selection of Type 1 AGN \citep{Richards06}.

%We note that there are a small number of sources that we spectroscopically classify as AGN that appear in the `star-forming' region of the mid-infrared colour-colour plots. These may be very obscured SF galaxies.% discussed in Section \ref{sc}. 

%RAY TO WRITE A PAR ABOUT HOW WE ARE DOING THIS FOR DEEPER RADIO DATA THAN BEFORE?

\begin{figure}
\begin{center}
\includegraphics[angle=0, scale=0.8]{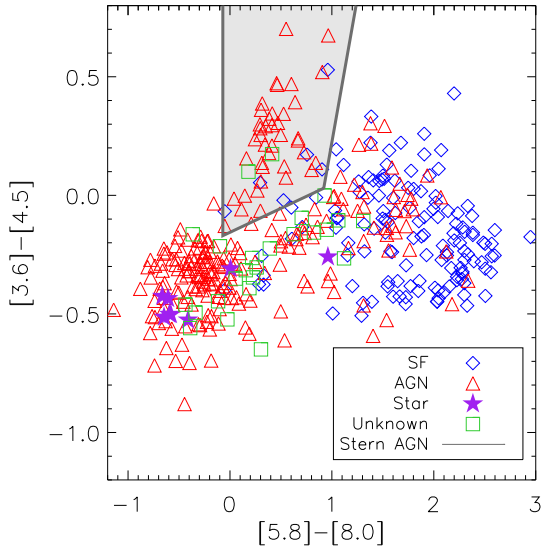}
\end{center}
\caption{MIR $[5.8\,\mu \rm m]-[8\,\mu$m] vs. $[3.6\,\mu \rm m]-[4.5\,\mu$m] colour-colour plot. The data in blue show the sources classified spectroscopically as SF and those in red show the sources classified as AGN. Stars are shown in purple and the sources we were unable to classify spectroscopically are shown in green. The grey shaded region shows the location of the \citet{Stern05} selection for AGN.}\label{ir_cols3}
\end{figure}

\citet{Stern05} used \emph{Spitzer} data from the AGN and Galaxy Evolution Survey (AGES) to identify Type 1 AGN and we find their selection method to be a more robust way of identifying broad-line AGNs. Of the 48 sources that satisfy the ``Stern Wedge'', only 6 had been spectroscopically classified as SF with the remaining 42 AGN comprising 25 ``AGNb'', 12 ``AGNe'', 3 ``E'' or ``E+OII'' and two ``UNKNOWN'' (Figure \ref{ir_cols3}).

\subsection{Mid-infrared Radio Correlation}
\begin{figure}
\begin{center}
\includegraphics[angle=0, scale=0.8]{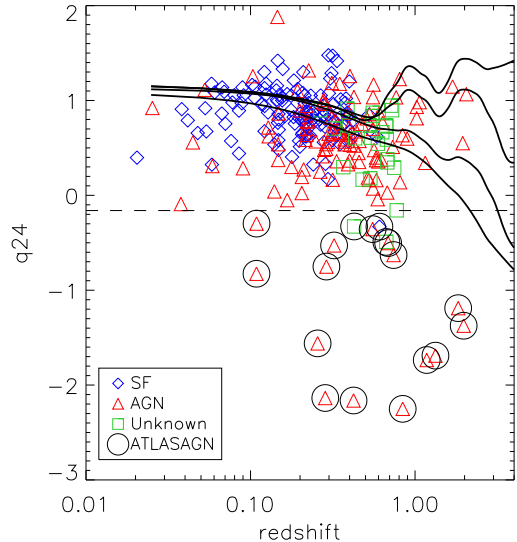}
\end{center}
\caption{The mid-infrared radio ratio, q$_{24}$, against redshift. The data in blue show the sources that were classified spectroscopically as SF and those in red show the sources that were classified as AGN. The sources which we were unable to classify spectroscopically are shown in green. The black lines are expected q$_{24}$ tracks derived from SED templates for galaxies with total infrared luminosities of 10$^9$ (normal galaxies), 10$^{11}$, 10$^{12}$ and 10$^{13} L_{\odot}$ (ULIRGs) going from bottom to top, from \citet{Chary01}. Sources with q$_{24}$ $\le -0.16$ are shown in black and the dotted line is at q$_{24}=-0.16$.}\label{frc24_2}
\end{figure}

\citet{Norris06} and \citet{Middelberg08} both use the Mid-Infrared Radio Correlation (MRC) to discriminate AGN from SF galaxies with ${\rm q}_{24}=\rm log(S_{24\mu \rm m}$/S$_{1.4 \rm GHz}) \le -0.16$ identified as AGN. This criterion ensures that sources have a ten-fold excess of radio emission over the MRC and are thus very likely AGN. We apply this discriminant and find 16 of the 466 sources satisfy this AGN criterion. The MRC using 24$\,\mu$m data is plotted against redshift in Figure \ref{frc24_2}.

%\begin{figure}
%\begin{center}
%\includegraphics[angle=0, scale=0.5]{../..//atlas_spectroscopy/SFvAGN/ircolours/FRC24_atlas_ourscompare.ps}
%\end{center}
%\caption{24$\,\mu$m flux density against 20\,cm flux density. The grey data are all sources in ATLAS for which IR data is available. The data in blue show the sources that were classified spectroscopically as SF and the data in red show the sources that were classified as AGN. Stars are shown in purple and the sources for which we were unable to classify spectroscopically are shown in green. Sources with q24 $\le$ -0.16 are shown in black.}\label{frc24_1}
%\end{figure}

Only one source presents a conflict between the spectroscopic classification and that based on q$_{24}$. The source, S385 in CDFS, is spectroscopically classified as a SF galaxy, and lies close to the selection boundary but, at z~=~0.6, it is difficult to assign a spectroscopic classification due to the dearth of strong optical features in the observed spectroscopic range.

%There is one source that we spectroscopically classified as 'SF?' and is classified as AGN according to the q24 diagnostic. The source, S385 has q24 = -0.33 and is at z = 0.6. It is entirely possible that the spectroscopic classification is incorrect however as the source is at a higher redshift and is not far from the q24 AGN cut-off we cannot unambiguously determine whether this source is primarily driven by SF or AGN mechanisms. %PAPER 2 WILL DO ALL THIS STUFF

The IR colours and the MRC in this paper are primarily used to determine the validity of the spectroscopic classifications that we presented in Section \ref{sc}. We hereby conclude that our spectroscopic classifications are broadly reliable and for the remainder of this paper we will use the spectroscopic classifications only. The 32 sources that we were unable to classify spectroscopically as SF or AGN could not be unambiguously classified using the mid-infrared diagnostics. 

We defer a more detailed exploration of classification limitations and ambiguities to a future paper that shall also utilise 843\,MHz and 2.3\,GHz ATLAS radio data from \citet{Randall12} and Zinn et al. (in preparation). We also note the strength of classification techniques such as the \emph{k}-Nearest Neigbour (\emph{k}NN) classifier, which can operate on multi-dimensional datasets \citep[e.g.][]{Gieseke11}. 

%will return to this in a future paper that shall discuss the various AGN/SF diagnostics in much greater detail. We shall use the spectroscopic indices from \citet{Randall12} and Zinn et al. (in preparation) to determine their relationship with the radio emission mechanism. Furthermore, we shall use these various diagnostics to explore other samples of objects \citep[e.g. the CSS sample of][]{Randall11}. We also note the strength of classification techniques such as the \emph{k}-Nearest Neigbour (\emph{k}NN) classifier, which can operate on multi-dimensional datasets \citep[e.g.][]{Gieseke11}. 

\subsection{Computing the Radio Luminosity Function}
The radio luminosity function (RLF) allows direct insight into the evolutionary history of galaxies. SF galaxies are known to positively evolve with redshift, that is, the radio luminosity and number density of SF galaxies is greater at higher redshifts \citep[][Dwelly et al., in preparation]{Hopkins98,Afonso05}. The evolution of powerful radio AGN is well understood with powerful radio AGN also evolving positively with redshift. The evolution of low-luminosity radio AGN is less well understood with some studies finding no evidence for any evolution of the RLF for low-luminosity radio AGN \citep[e.g.][]{Clewley04} and others finding that low-luminosity AGN do evolve with redshift, albeit more slowly than their high-luminosity counterparts \citep[e.g][]{Smolcic09,McAlpine11}. Deep radio data are being used to study low-luminosity AGN \citep[e.g][]{Kimball11,Padovani11}. \citet{Best12} suggest that the luminosity dependence of the evolution of the AGN RLF may be attributed to the varying fractions of `hot' and `cold' mode sources with redshift. 

In order to calculate luminosity functions, the sample must have known redshifts as well as both an optical magnitude limit and a radio flux density limit \citep{Schmidt68,Schmidt77}. Moreover, the sample volume must be large enough to mitigate the effects of cosmic variance \citep{Moster11}. It is imperative that any incompleteness in the dataset is well understood. 

%For the calculation of the RLF, all redshift information and spectroscopic classifications come from the spectroscopic data presented in this paper. 

\subsubsection{RLF Survey Area}
The ATLAS radio data cover $\sim 7.5$~square degrees on the sky, 2.96~square degrees in CDFS and 4.69~square degrees in ELAIS. AAOmega observations of ATLAS comprised two overlapping pointings in CDFS and a single pointing in ELAIS. The total area covered by AAOmega in CDFS is 2.83~square degrees and the total area covered by AAOmega in ELAIS is $\pi$~square degrees. The discrepancies in size and coverage are due to CDFS's shape and the AAOmega observations actually extended beyond the boundary of the radio observations, whereas in ELAIS the AAOmega pointing was within the radio observations. For the construction of the RLF, the survey area must be limited to where there is data at both radio and optical bands. 

Due to the presence of a strong ($>1$~Jy) source near the centre of the CDFS field \citep{Norris06} the CDFS radio image contains a number of artefacts. More advanced data reduction techniques have been applied to DR2 and DR3 and the CDFS radio image from these data releases is less affected by artefacts and has a more uniform rms. Due to the varying rms of the DR1 CDFS radio image, the remainder of this Section will deal only with ELAIS\footnote{Although we could have taken the subset of the CDFS radio image that was uniform, this area has fewer than 30 radio sources with optical counterparts, so we construct the RLF for ELAIS only.}. The RLF for CDFS will be deferred to the release of DR2 and DR3. 

The AAT observations for ELAIS cover only a single pointing so we limit the total area with which we can construct the RLF to $\pi$~square degrees, the field-of-view (FOV) of AAOmega.

%Unfortunately, while this method was successfully applied to ELAIS, we were unable to approximate the CDFS radio image to a number of subimages with uniform sensitivities. We attribute this to the image artefacts caused by the very strong source in the CDFS field. More advanced data reduction techniques have been applied to DR2 and DR3 and the CDFS radio image from these data releases is much clearer of artefacts and has a more uniform rms. Due to the varying rms of the DR1 CDFS radio image, the remainder of this Section will deal only with ELAIS\footnote{Although we could have taken the subset of the CDFS radio image that was uniform, this area has $<$30 radio sources with optical counterparts so we construct the RLF for ELAIS only.}. The RLF for CDFS will be presented upon the release of DR2 and DR3. 

\subsubsection{The Radio Data}
\begin{figure}
\begin{center}
  \includegraphics[angle=0, scale=0.8]{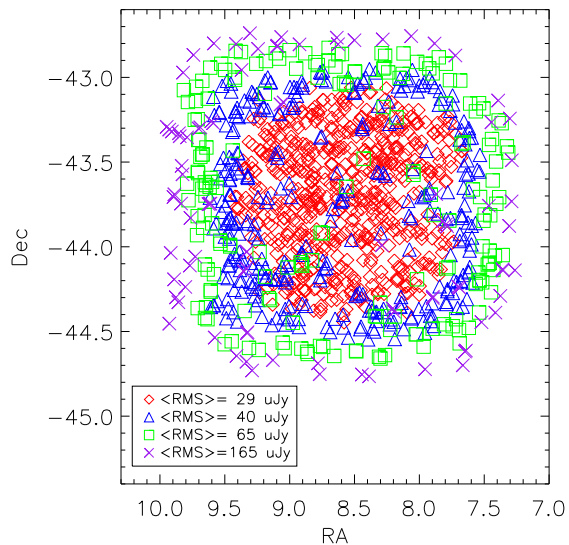}
\end{center}
\caption{ELAIS's rms characteristics, with four different rms bins chosen, and colour-coded.}\label{elaisrms}
\end{figure}

\begin{figure}
\begin{center}
  \includegraphics[angle=0, scale=0.8]{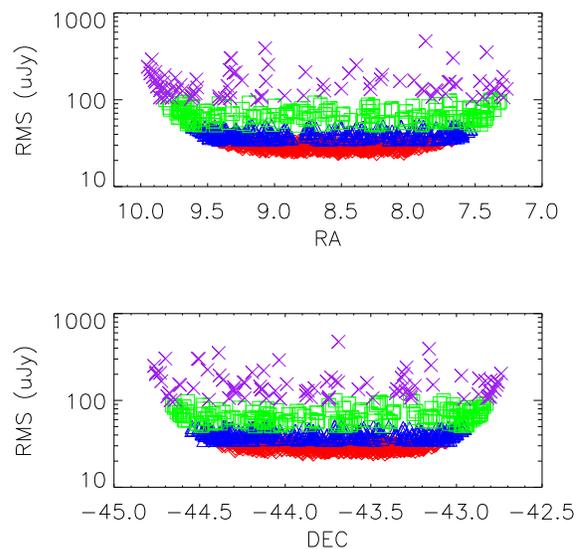}
\end{center}
\caption{RMS against both RA and Dec for ELAIS with the same symbol scheme as above.}\label{elaisradeccut}
\end{figure}

ATLAS DR1 is nominally quoted as having an rms of 30\,$\mu$Jy\,beam$^{-1}$, but the sensitivity of the radio images is not uniform, typically being less sensitive closer to the edges. Figures \ref{elaisrms} and \ref{elaisradeccut} shows the varying rms in ELAIS at each radio source position. These values were calculated by producing an rms map and calculating the mean rms in the 75 $\times$ 75 pixels ($\sim$15 $\times$ 15 synthesised beams) surrounding the source position. To mitigate the effect of outlier pixels the brightest and faintest pixel in the rms grid were discarded when calculating the mean rms.

%to \ref{cdfsradeccut}

%CDFS has the additional issue of a strong ($>$1Jy) source near the centre of the field. 

%\begin{figure}
%\begin{center}
%  \includegraphics[angle=0, scale=0.8]{/Users/mao006/phd/atlas_spectroscopy/rmsmaps/cdfs_radec_rms.ps}
%\end{center}
%\caption{RA-Dec plot of CDFS's rms characteristics, with four different rms bins chosen, and colour-coded.}\label{cdfsrms}
%\end{figure}

%\begin{figure}
%\begin{center}
%  \includegraphics[angle=0, scale=0.8]{/Users/mao006/phd/atlas_spectroscopy/rmsmaps/cdfs_rms_slice.ps}
%\end{center}
%\caption{RMS plotted against both RA and Dec for CDFS with the same symbol scheme as above.}\label{cdfsradeccut}
%\end{figure}

In order to construct the radio luminosity function we must calculate the maximum volume to which a source of a given radio luminosity may be detected (V$_{\rm MAX}$). If the radio image had uniform sensitivity, 

\begin{eqnarray}\label{vmaxeqn}
\rm V_{MAX}&=&\frac{4}{3}\pi\Omega\rm D_{MMAX}^3
\end{eqnarray}
where
$\Omega$ is the solid angle (in steradians) subtended by the image, and
D$_{\rm MMAX}$ is the maximum comoving distance in Mpc for a source of a given radio luminosity \citep{Hogg99}.

In order to account for the non-uniform sensitivity however, we must treat the radio image as a number of nested subimages, each with its own limiting flux density. Consequently, each subimage may be approximated by a uniform radio image, and sources that fall below $5\times$ rms (Figure \ref{elaisrms}) are discarded. For a sample with regions that have different sensitivies, the maximum volume available to each source may be calculated using the method described in section IIc of \citet{Avni80}. 

 Using the Miriad task \textsc{imhist} we are able to calculate the total area of the radio image above each chosen flux limit. Consequently, for each source we are able to calculate the total volume available to the source, as well as the actual volume enclosed by that source using the method of \citet{Avni80}. The calculations for V$_{enclosed}$ and V$_{available}$ are shown in Appendix \ref{Vmaxradio}. %Next we turn to the optical completeness corrections. %The dataset is however not yet complete enough to derive the RLF as we also need to account for the optical magnitude limit. 

%\begin{table}
%\begin{center}
%\caption{MAY REMOVE THIS - JUST FOR NOW...}\label{rms}
%\begin{tabular}{crrrrr}
%\\
%\hline
%RMS limits& No.  & $<$RMS$>$\\
%($\mu$Jy\,beam$^{-1}$) & & ($\mu$Jy\,beam$^{-1}$)\\
%\hline
%\\
%ELAIS\\
%$<$35 & 166 & 29 $\pm$ 3\\
%35 $\le$ RMS $<$50 & 87 & 40 $\pm$ 4\\
%50 $\le$ RMS $<$100 & 44 & 65 $\pm$ 10\\
%100 $\le$ RMS & 9 & 165 $\pm$ 53\\
%\\
%CDFS\\
%$<$35 & 52 & 28 $\pm$ 5\\
%35 $\le$ RMS $<$50 & 53 & 41 $\pm$ 4\\
%50 $\le$ RMS $<$100 & 45 & 65 $\pm$ 13\\
%100 $\le$ RMS & 10 & 389 $\pm$ 507\\
%\hline
%\end{tabular}
%\end{center}
%\end{table}

%From our AAT observations we were able to successfully determine redshifts for 466 of the ATLAS sources. 

\subsubsection{The Optical Data}
We use the SuperCOSMOS \citep{Hambly01} optical magnitude data, which is limited at $B_j$~$\sim$~23 and $R\sim$~22. Sources fainter than these limits are discarded. Although deeper photometry exist within parts of the ATLAS fields \citep[e.g.,][]{Lonsdale03}, SuperCOSMOS is the only survey to date that covers our region of interest uniformly. Moreover, the faintest sources observed successfully in our spectroscopic observations using the AAT have $R\sim 22$. Optical $k$-corrections were applied using the method from \citet{DePropris04}. Because both $R$ and $B$ band magnitudes are necessary to calculate the $k$-correction when calculating absolute magnitudes, we must determine magnitude limits for our spectroscopic observations for both photometric bands. 

\begin{figure*}
\begin{center}
\begin{tabular}{cc}
\includegraphics[angle=0, scale=0.8]{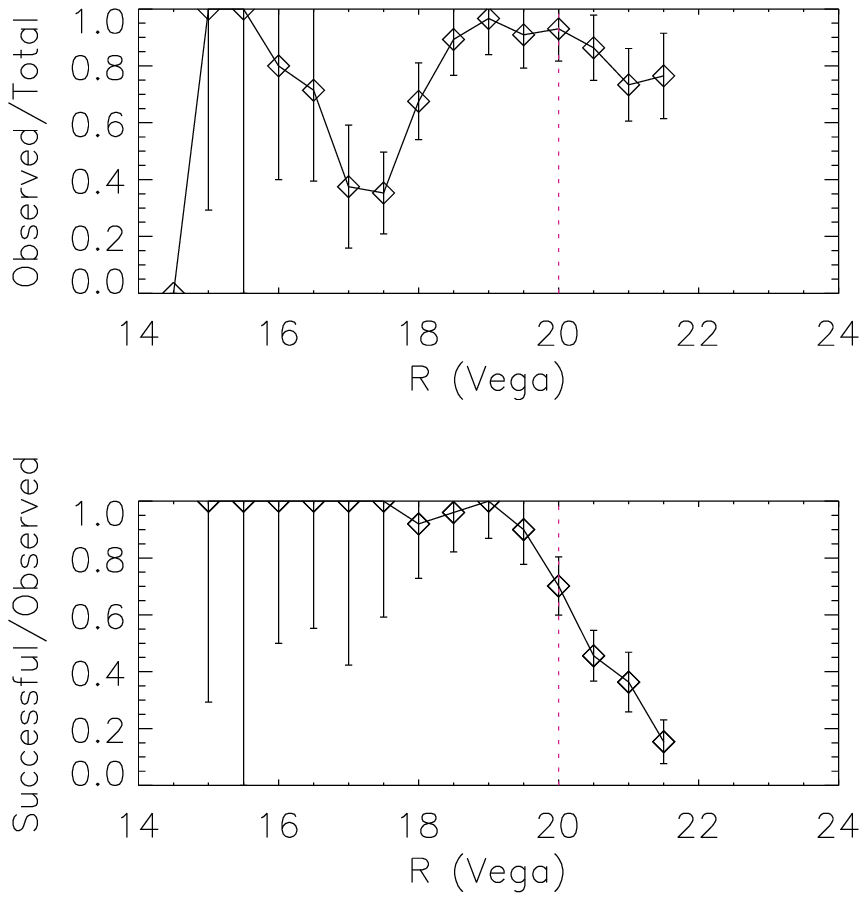} & \includegraphics[angle=0, scale=0.8]{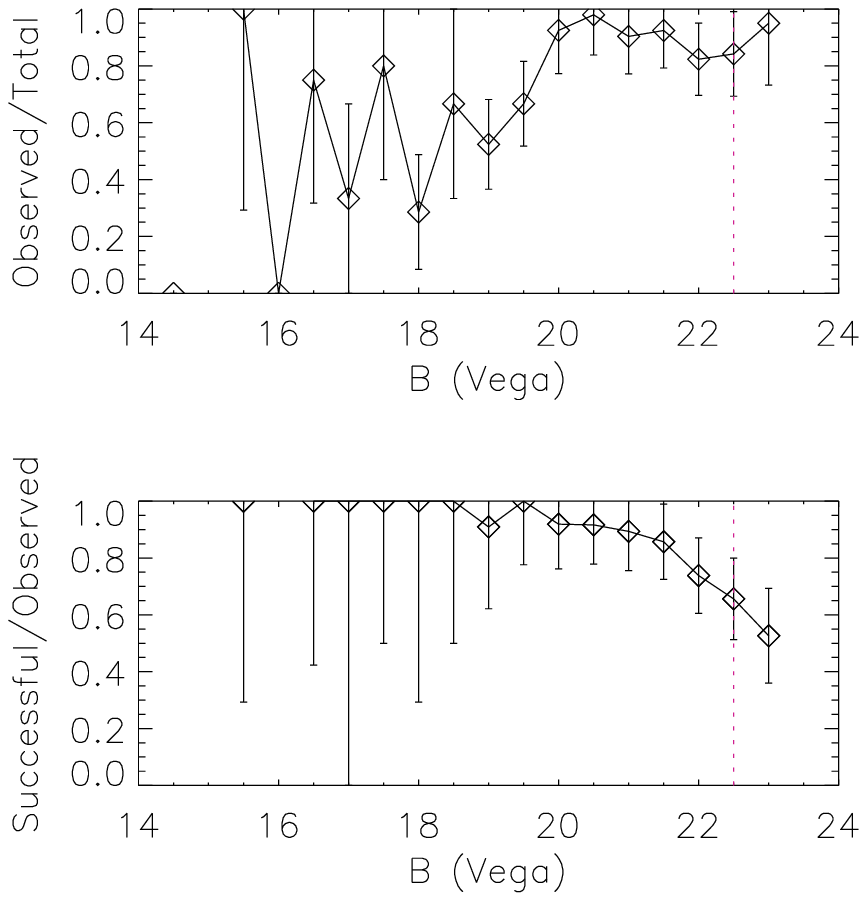}\\
\end{tabular}
\end{center}
\caption{The top panels shows the fraction of sources observed against optical magnitude (left: $R$-band, right: $B$-band) within the FOV and the bottom panel shows the fraction of sources whose observations yielded quality spectra. The pink dotted lines indicate the magnitude limits. The error bars are Poisson. The error bars at bright magnitudes are large due to the small numbers in each bin. }\label{complete_r2}
\end{figure*}

Figure \ref{complete_r2} shows the completeness functions for both bands. We choose $R_{lim}$=20 and $B_{lim}$=22.5. There is still some incompleteness ($\sim$30 per cent) at these magnitude limits, but we choose these limits to maximise the number of sources that we can use to calculate the luminosity function. There is an additional incompleteness that we must account for (top panels of Figure \ref{complete_r2}) due to the fraction of sources that were not observed with the AAT. This is, in part, due to the poor observing conditions. In order to observe a dense field of sources completely it is often necessary to observe a number of fibre configurations and the poor observing conditions limited our ability to observe multiple configurations. We also selected against bright sources ($R<18$) because the magnitude range for any single observation is limited to $\sim$2 magnitudes to avoid the problem of scattered light. To account for these completeness issues, we take the product of these two corrections (C$_i$) and weight each source of a given magnitude by this correction when calculating the RLF (below). We can now calculate the maximum volume out to which the optical source could be detected. As two optical bands are necessary for $k$-correcting this dataset, we calculate V$_{\rm MAX}$ at both bands and take the smaller of the two to be the limiting value. The maximum volume to which the optical source can be detected may be calulated using Equation \ref{vmaxeqn}. 

% RGS and MM attempted to measure spectra for a number of these `bright' sources at a later date but found many of these had large angular sizes and were probably very low redshift SF galaxies.

%\begin{figure}
%\begin{center}
%\includegraphics[angle=0, scale=0.8]{../..//atlas_spectroscopy/lumfnNEW/completeness/completeness_bmag20mag_elais.ps}
%\end{center}
%\caption{The top panel shows the fraction of sources observed against $B$-band magnitude within the FOV and the bottom panel shows the fraction of sources whose observations yielded quality spectra. The error bars are Poissonion errors. The error bars at bright magnitudes are large due to the small numbers in each bin.}\label{complete_bj}
%\end{figure}

\subsubsection{V/V$_{\rm MAX}$ test}
All sources fainter than the radio flux density and optical magnitude limits for our dataset have now been discarded. We remove any sources with stellar spectra as these are believed to be chance alignments, leaving a total of 226 sources in ELAIS out of the original 306 with which to calculate the radio luminosity function. Perhaps unsurprisingly, 71 per cent (17/24) of the `unknown' spectroscopic-type sources were discarded due to their faint optical magnitudes. We do not account for the `unknown' spectra for the remainder of this section. Although they will undoubtedly affect either or both the SF or the AGN RLF, there are only seven of these sources in our sample and their inclusion would not significantly change any of our results. 

We can now check the completeness of our dataset using the V/V$_{\rm MAX}$ test of \citet{Schmidt68, Schmidt77}. We calculated both the available and enclosed volumes for each radio source, as well as both the volume and maximum volume out to which the optical source might be detected. We compare the maximum volume out to which the optical source may be detected (Equation \ref{vmaxeqn}) with the maximum volume available to each radio source (Equation \ref{Vavailableeqn}) and take the smaller of the two as the limiting value, as in Equation \ref{whichvmax}. 

\begin{eqnarray}\label{whichvmax}
\rm V_{MAX}&=&{\rm min}[\rm V_{(\rm MAX, OPT)}, V_{(\rm available, RAD)}]
\end{eqnarray}
where
V$_{\rm MAX,OPT}$ is the maximum volume to which the optical source may be detected, and 
V$_{\rm available, RAD}$ is the maximum volume available to the radio source. 

Of the 226 radio sources in ELAIS with which we are able to construct the RLF, 146 (65 per cent) are detection-limited by the radio flux density limits. The detection of sources with radio luminosities greater than 10$^{24}$~W~Hz$^{-1}$ is almost entirely governed by the optical magnitude limits. There is no discernable difference in the detection limit between SF and AGN. We calculate $\langle$V/V$_{\rm MAX}\rangle$ to be 0.563 $\pm$ 0.011, which is consistent with positive evolution as $\langle$V/V$_{\rm MAX} \rangle >0.5$.%, which is sufficiently close to 0.5 thus we deem our sample to be complete for the purposes of constructing the radio luminosity function.

%Figure \ref{dlflag} demonstrates that t
%\begin{figure}
%\begin{center}
%\includegraphics[angle=0, scale=0.5]{../..//atlas_spectroscopy/lumfnNEW/dlflag.ps}
%\end{center}
%\caption{The fraction of sources whose detection is limited by the optical data plotted against radio luminosity. DOES THIS PLOT ADD ANYTHING? }\label{dlflag}
%\end{figure}

The radio luminosity function for our sources was calculated using

\begin{equation}
\Phi = \displaystyle\sum\limits_{i=0}^n\frac{1}{{\rm V}_{{\rm MAX},i}} \pm \sqrt{\displaystyle\sum\limits_{i=0}^n\frac{1}{({\rm V}_{{\rm MAX},i})^2}}
\end{equation}

where
$\Phi$ is the density of sources in Mpc$^{-3}$dex$^{-1}$,
$n$ is the number of sources, and
V$_{\rm MAX}$ is the maximum volume out to which we can see the source. 

%We have also scaled $\Phi$ with the bin size so it is measured in Mpc$^{-3}$dex$^{-1}$. 

To correct for incompleteness, we apply the corrections from Figure \ref{complete_r2}. The optical band that limits V$_{\rm MAX}$ is used to determine which completeness correction is used. 

\begin{equation}
\Phi = \displaystyle\sum\limits_{i=0}^n\frac{1}{{\rm C}_i \times {\rm V}_{{\rm MAX},i}} \pm  \displaystyle\sum\limits_{i=0}^n\frac{1}{\sqrt(\Delta {\rm V}_{{\rm MAX},i}^2+\Delta {\rm C}_i^2)}
\end{equation}
where
C$_i$ is from the completeness functions (Figure \ref{complete_r2}). Where the number of sources is small ($n<40$), the standard $\sqrt{n}$ approximation to Poisson errors are inaccurate, so for $n<40$ we have evaluated the 1-sigma Poisson errors using \citet{Gehrels86}. 

%The SF and AGN RLFs appear to cross at $\sim$10$^{22.9}$ W$\,$Hz$^{-1}$. BUT KINDA DEPENDS ON BINSIZE

\subsection{The RLF for ELAIS}
\begin{figure}
\begin{center}
\includegraphics[angle=0, scale=0.8]{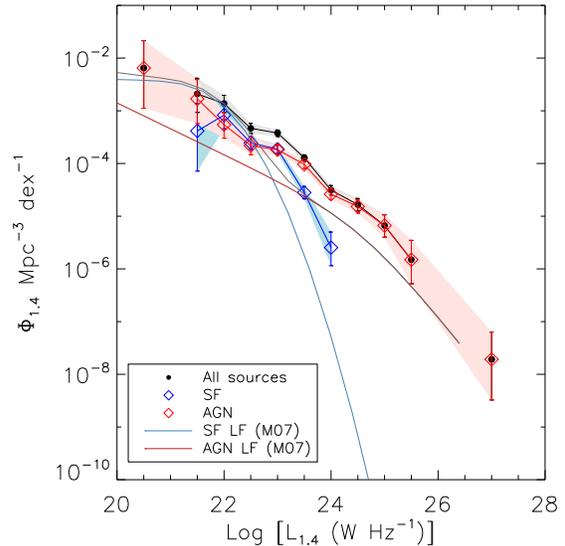}
\end{center}
\caption{The blue diamonds are the RLF for SF sources, the red diamonds are for the AGN RLF and black is the total. The solid pale blue and pink lines are the parametric fits to the local RLF from \citet{Mauch07}. Errors are Poisson.}\label{lumfn_SFvAGN}
\end{figure}

Figure \ref{lumfn_SFvAGN} and Table \ref{totalrlftab} show the resulting RLF over our entire redshift range. We overplot data from \citet{Mauch07} who calculated the local radio luminosity functions for both SF galaxies and radio-loud AGN. The data from \citet{Mauch07} span a redshift range of $0.003 <$~z~$< 0.3$ with a median redshift of 0.043. The SF galaxy RLF for ATLAS-ELAIS appears to be higher than the SF galaxy RLF of \citet{Mauch07}, especially for more luminous sources. This may be attributed to our more sensitive radio observations, which results in a higher median redshift for our sample. The median redshift for our complete sample is z~$\sim$~0.316 compared to \citet{Mauch07} whose median sample redshift is z~$\sim$~0.043. The SF galaxy RLF is known to positively evolve with redshift, so our higher SF RLF is likely due to cosmic evolution of the SF RLF. The AGN RLF also appears to be higher than the AGN RLF of \citet{Mauch07} by a factor of $\sim$3. This may be due to evolution, cosmic variance, differences between how we classify sources, particularly composite sources, or a combination of these factors. Detailed discussion of the SF galaxy and AGN luminosity functions are presented below in Sections \ref{sfrlf} and \ref{agnrlf}. 

\renewcommand{\arraystretch}{1.3}

\begin{table*}
\begin{center}
\caption{Radio luminosity function at 1.4\,GHz for all radio sources, SF galaxies and AGN in ELAIS.}\label{totalrlftab}
\begin{tabular}{lrcrcrc}
\\
\hline
& & Total & & SF & & AGN\\
log$_{10}$L$_{1.4}$ & N & log$_{10}\Phi$ & N & log$_{10}\Phi$& N & log$_{10}\Phi$\\
(W~Hz$^{-1}$) & & (Mpc$^{-3}$~dex$^{-1}$) & & (Mpc$^{-3}$~dex$^{-1}$) & & (Mpc$^{-3}$~dex$^{-1}$)\\
\hline
        20.5&           1&       -2.19$^{+        0.52}_{       -0.76}$&           0&        &           1&       -2.19$_{       -0.76}^{+        0.52}$\\
        21.0&           0&        &           0&        &           0&        \\
        21.5&           3&       -2.68$^{+        0.30}_{       -0.35}$&           1&       -3.38$_{       -0.76}^{+        0.52}$&           2&       -2.77$_{       -0.47}^{+        0.37}$\\
        22.0&          10&       -2.86$^{+        0.16}_{       -0.17}$&           5&       -3.08$_{       -0.26}^{+        0.23}$&           5&       -3.26$_{       -0.26}^{+        0.23}$\\
        22.5&          25&       -3.33$^{+        0.10}_{       -0.10}$&          16&       -3.61$_{       -0.12}^{+        0.12}$&           9&       -3.66$_{       -0.18}^{+        0.16}$\\
        23.0&          67&       -3.42$^{+        0.06}_{       -0.07}$&          32&       -3.73$_{       -0.08}^{+        0.08}$&          33&       -3.74$_{       -0.08}^{+        0.08}$\\
        23.5&          66&       -3.89$^{+        0.06}_{       -0.07}$&          17&       -4.55$_{       -0.12}^{+        0.12}$&          48&       -4.02$_{       -0.08}^{+        0.07}$\\
        24.0&          28&       -4.51$^{+        0.09}_{       -0.09}$&           3&       -5.60$_{       -0.34}^{+        0.30}$&          22&       -4.58$_{       -0.10}^{+        0.10}$\\
        24.5&          17&       -4.78$^{+        0.12}_{       -0.12}$&           0&        &          16&       -4.82$_{       -0.12}^{+        0.12}$\\
        25.0&           6&       -5.18$^{+        0.20}_{       -0.22}$&           0&        &           6&       -5.18$_{       -0.22}^{+        0.20}$\\
        25.5&           2&       -5.82$^{+        0.37}_{       -0.46}$&           0&        &           2&       -5.82$_{       -0.46}^{+        0.37}$\\
        26.0&           0&        &           0&        &           0&        \\
        26.5&           0&        &           0&        &           0&        \\
        27.0&           1&       -7.72$^{+        0.52}_{       -0.77}$&           0&        &           1&       -7.72$_{       -0.77}^{+        0.52}$\\
Total&         226&   &          74&&         145&\\
$<$V/V$_{\rm MAX}>$&&       0.564$\pm$       0.011&&       0.594$\pm$       0.020&&       0.532$\pm$       0.013\\
\hline
\end{tabular}
\end{center}
\end{table*}
\renewcommand{\arraystretch}{1.0}

%In order to probe the cosmic evolution of SF galaxies and AGN, we must construct the RLF for different redshift slices.% Most of the higher redshift sources are removed when we take into account the optical and radio limits of our observations. In fact,  $<$10 per cent (22/226) of the sources in our sample are at z~$>$~0.5. Consequently, we probe the evolution of the luminosity function of ATLAS sources to z~=~0.5 as redshift bins above this will be subject to small number statistics.

% (Figure \ref{zhist_haslum})

%\begin{figure}
%\begin{center}
%\includegraphics[angle=0, scale=0.5]{../..//atlas_spectroscopy/lumfnNEW/elais_haslum_zhist.ps}
%\end{center}
%\caption{Histogram of redshifts of the radio flux density and optical magnitude limited sample. There are 226 sources in the histogram. SF galaxies are shown in blue and AGN are shown in red. The black line shows the overall histogram and the bin width is 0.1z.}\label{zhist_haslum}
%\end{figure}

\begin{figure}
\begin{center}
\includegraphics[angle=0, scale=0.8]{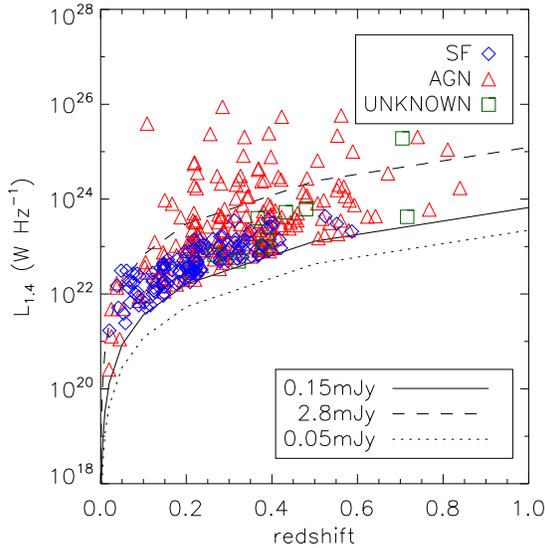}
\end{center}
\caption{Radio luminosity plotted against redshift for our sources. The lines show different flux density limits. The solid line corresponds to a flux limit of 150\,$\mu$Jy\,beam$^{-1}$, which is $\sim$5~$\times$~rms for our data. The dashed line corresponds to the flux density limit for \citet{Mauch07} and the dotted line corresponds to the flux density limit for COSMOS \citep{Smolcic09}. }\label{lumcompare}
\end{figure}

To demonstrate the completeness of our radio data, Figure \ref{lumcompare} plots the radio luminosity against redshift with flux limits overplotted. For example, in ATLAS, radio sources with powers of 10$^{23}$\,W~Hz$^{-1}$ may be detected to z $\sim$ 0.4. The dashed line in Figure \ref{lumcompare} shows the flux density limit for the study from \citet{Mauch07}. The dotted line in Figure \ref{lumcompare} shows the flux density limit for the COSMOS survey \citep{Scoville07}. \citet{Smolcic09} calculated the RLF for low-power AGN in the COSMOS sample using predominately photometric redshifts and found mild evolution to z$\sim$1.3. 

% The ATLAS data radio data is significantly deeper than that used by \citet{Mauch07} and can probe the evolution of low luminosity sources to higher redshifts.
% While COSMOS has deeper radio data than ATLAS, their optical data is less complete than ATLAS's and hence the redshifts and spectroscopic classifications for ATLAS are more complete.

\subsection{RLF for SF galaxies}\label{sfrlf}
\begin{figure}
\begin{center}
\includegraphics[angle=0, scale=0.8]{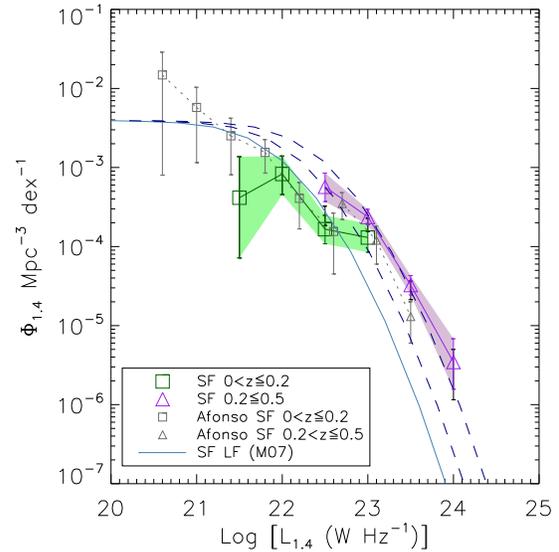}
\end{center}
\caption{Radio luminosity function for SF galaxies in ATLAS. We have split our data into two redshift bins so as to allow for direct comparison against \citet{Afonso05}. The grey data points are from \citet{Afonso05}. Errors are Poisson. The parametric fit to the SF RLF from \citet{Mauch07} is also shown, both in its original form and after applying luminosity evolution in the form L$_{1.4\,GHz} \propto$~(1+z)$^{Q}$, with $Q=2.7\pm0.6$ \citep{Hopkins04} and z~=~0.2 and z~=~0.5.}\label{sflfcompare}
\end{figure}

The RLF for SF galaxies has been proposed to evolve positively with redshift \citep[e.g.][]{Rowanrobinson93,Hopkins98}. We compare our RLF for SF galaxies with that of previous work by \citet{Afonso05} and \citet{Mauch07} (Figure \ref{sflfcompare} and Table \ref{rlfevolvetab}). Although the total SF galaxy RLF for our dataset is higher than the SF galaxy RLF of \citet{Mauch07} -- especially for more luminous sources -- this is attributed to cosmic evolution. When the SF galaxy RLF is separated into two redshift bins\footnote{In order to construct the RLF for different redshift slices it is necessary to impose a maximum and minimum z. Consequently, V$_{\rm MAX}$ = min(V$_{\rm ZMAX}, V_{\rm MAX})-$~V$_{\rm ZMIN}$.} it is apparent that the low redshift bin is in good agreement with \citet{Mauch07}. \citet{Afonso05} found their data were consistent with pure luminosity evolution in the form L$_{1.4\,GHz} \propto$~(1+z)$^{Q}$, with $Q=2.7\pm0.6$ \citep{Hopkins04}. Our dataset has very similar optical and radio limits to \citet{Afonso05}, and we find our SF RLF to agree within the uncertainties with theirs at both redshift bins. %The fact that the luminosity and number density of SF galaxies is greater at higher redshifts is consistent with downsizing \citep{Cowie96}. 

\renewcommand{\arraystretch}{1.3}
\begin{table*}
\begin{center}
\caption{Radio luminosity function at 1.4\,GHz for SF galaxies and AGN in ELAIS at two different redshift bins.}\label{rlfevolvetab}
\tiny{
\begin{tabular}{lrcrcrcrcrc}
\\
\hline
& & SF 0$<$z$\le$0.2 & & SF 0.2$<$z$\le$0.5& & AGN 0$<$z$\le$0.2& & AGN 0.2$<$z$\le$0.4& & AGN 0.4$<$z$\le$0.8\\
log$_{10}$L$_{1.4}$ & N & log$_{10}\Phi$ & N & log$_{10}\Phi$& N & log$_{10}\Phi$& N & log$_{10}\Phi$& N & log$_{10}\Phi$\\
(W~Hz$^{-1}$) & & (Mpc$^{-3}$~dex$^{-1}$) & & (Mpc$^{-3}$~dex$^{-1}$) & & (Mpc$^{-3}$~dex$^{-1}$) & & (Mpc$^{-3}$~dex$^{-1}$)& & (Mpc$^{-3}$~dex$^{-1}$)\\
\hline
        20.5&           0&   &           0&   &           1&       -2.19$_{       -0.76}^{+        0.52}$&           0&   &           0&   \\
        21.0&           0&   &           0&   &           0&   &           0&   &           0&   \\
        21.5&           1&       -3.38$_{       -0.76}^{+        0.52}$&           0&   &           2&       -2.77$_{       -0.47}^{+        0.37}$&           0&   &           0&   \\
        22.0&           5&       -3.08$_{       -0.26}^{+        0.23}$&           0&   &           5&       -3.26$_{       -0.26}^{+        0.23}$&           0&   &           0&   \\
        22.5&           8&       -3.78$_{       -0.19}^{+        0.17}$&           8&       -3.25$_{       -0.19}^{+        0.17}$&           6&       -3.73$_{       -0.23}^{+        0.20}$&           3&       -3.62$_{       -0.34}^{+        0.30}$&           0&   \\
        23.0&           8&       -3.89$_{       -0.19}^{+        0.17}$&          24&       -3.63$_{       -0.10}^{+        0.10}$&           5&       -3.90$_{       -0.26}^{+        0.23}$&          27&       -3.64$_{       -0.09}^{+        0.09}$&           1&       -5.03$_{       -0.76}^{+        0.52}$\\
        23.5&           0&   &          16&       -4.49$_{       -0.12}^{+        0.12}$&           2&       -4.35$_{       -0.49}^{+        0.37}$&          35&       -3.97$_{       -0.08}^{+        0.08}$&          11&       -4.43$_{       -0.15}^{+        0.15}$\\
        24.0&           0&   &           3&       -5.46$_{       -0.34}^{+        0.30}$&           1&       -4.49$_{       -0.83}^{+        0.52}$&          11&       -4.53$_{       -0.15}^{+        0.15}$&          10&       -4.95$_{       -0.16}^{+        0.15}$\\
        24.5&           0&   &           0&   &           0&   &           7&       -4.71$_{       -0.20}^{+        0.19}$&           7&       -4.98$_{       -0.20}^{+        0.19}$\\
        25.0&           0&   &           0&   &           0&   &           5&       -4.87$_{       -0.25}^{+        0.22}$&           1&       -6.10$_{       -0.77}^{+        0.52}$\\
        25.5&           0&   &           0&   &           0&   &           1&       -5.69$_{       -0.77}^{+        0.52}$&           1&       -5.83$_{       -0.77}^{+        0.52}$\\
Total&          22&&          51&&          22&&          89&&          31\\
$<$V/V$_{\rm MAX}>$&&       0.632$\pm$       0.039&&       0.480$\pm$       0.019&&       0.526$\pm$       0.032&&       0.548$\pm$       0.017&&       0.578$\pm$       0.030\\
\hline
\end{tabular}
}
\end{center}
\end{table*}

\renewcommand{\arraystretch}{1.0}

\subsection{RLF For AGN}\label{agnrlf}
%It is well established that powerful AGN ($>$10$^{26}$~W~Hz$^{-1}$) undergo rapid evolution \citep{Dunlop90}, and lower-luminosity AGN appear to evolve less rapidly than the most luminous sources due to the distribution of radio source counts \citep{Longair66}. While some studies find that low-luminosity AGN undergo no evolution \citep[e.g.][]{Clewley04}, \citet{Sadler07} were able to show that low-luminosity AGN (10$^{24}$ to 10$^{25}$~W~Hz$^{-1}$) undergo significant evolution over the redshift range 0$<$z$<$0.7.

\begin{figure}
\begin{center}
\includegraphics[angle=0, scale=0.8]{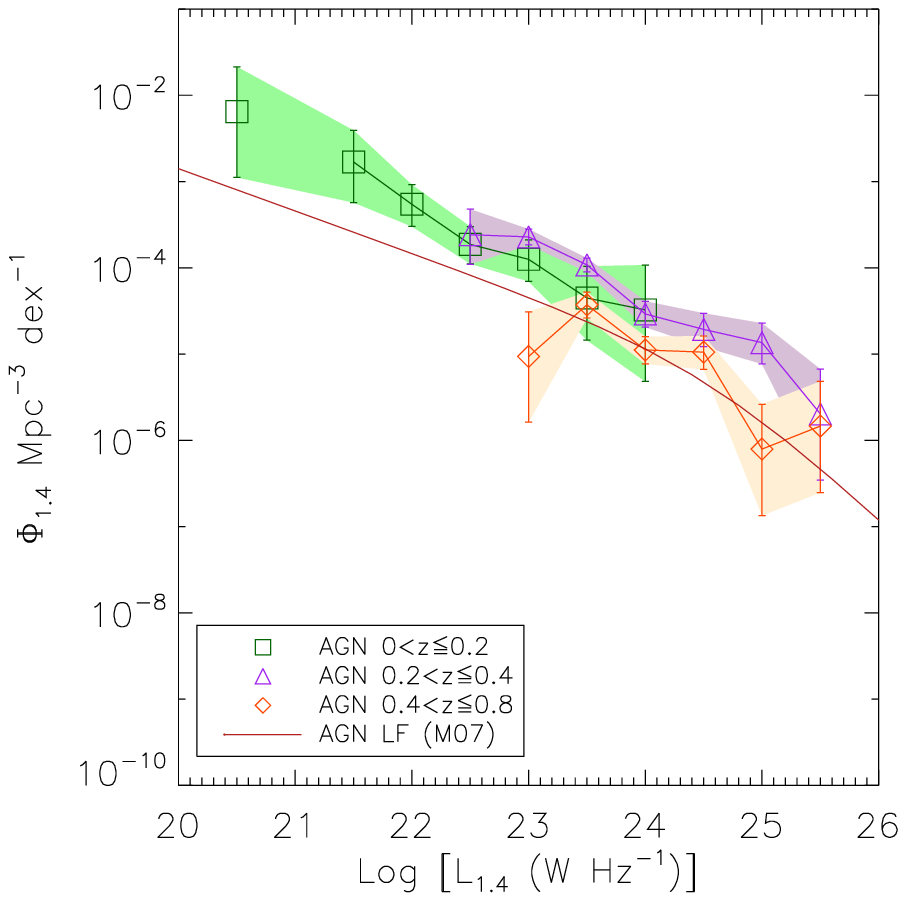}
\end{center}
\caption{Radio luminosity function for AGN in ATLAS-ELAIS. We have split our data into three redshift bins. The pink line is the local AGN RLF from \citet{Mauch07}. Errors are Poisson.}\label{agnlf}
\end{figure}

To determine the cosmic evolution of the RLF for AGN, we bin the data in redshift and compute the RLFs for each redshift bin. Figure \ref{agnlf} and Table \ref{rlfevolvetab} show the RLFs for the three redshift bins. We compare our AGN RLF with \citet{Mauch07} and find that our data is systematically inconsistent with the AGN RLF from \citet{Mauch07}, being a factor of $\sim$3 higher. 

\begin{figure}
\begin{center}
\includegraphics[angle=0, scale=0.8]{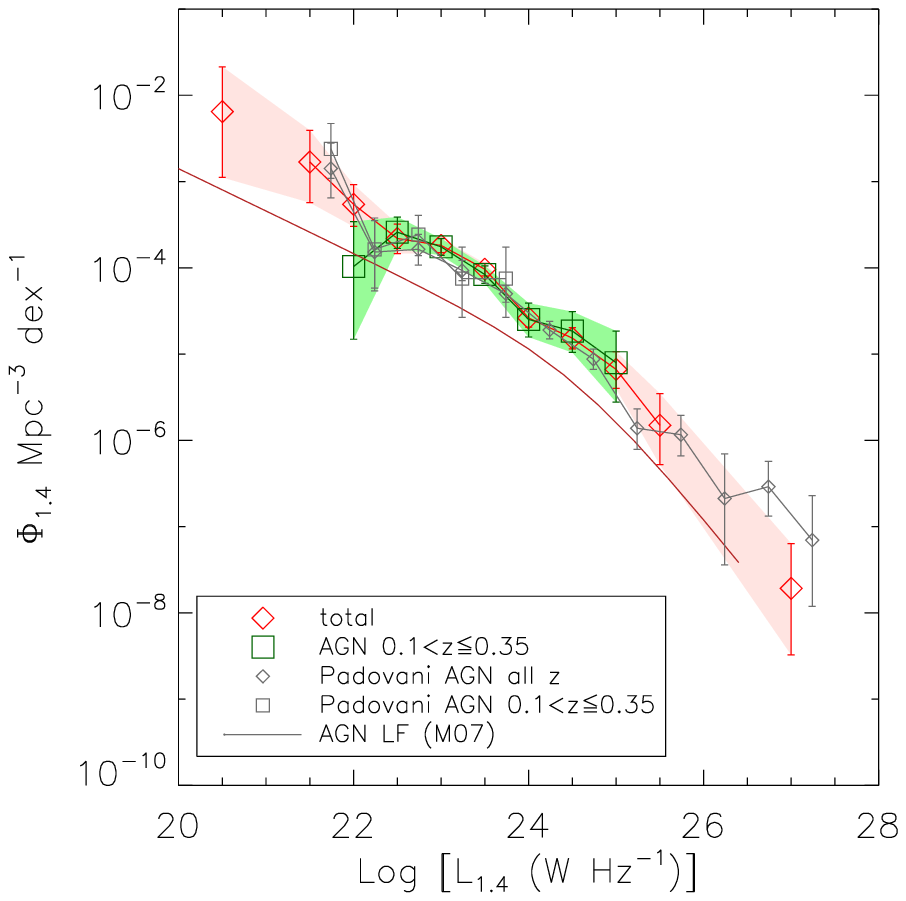}
\end{center}
\caption{Radio luminosity function for AGN in ATLAS-ELAIS with data from \citet{Padovani11} overplotted. Both the total AGN RLF and the RLF for the redshift range $0.1<$~z~$\le 0.35$ are plotted to allow direct comparison with \citet{Padovani11}. The pink line is the local AGN RLF from \citet{Mauch07}. Errors are Poisson.}\label{agnlfcfpadovani}
\end{figure}

\citet{Padovani11} calculated the RLF for AGNs using the VLA-CDFS sample \citep{Kellermann08} and also find their RLF to be a factor of 3 -- 4 higher than \citet{Mauch07}, predominately due to the population of ``radio-quiet'' AGNs that cannot be detected in less sensitive surveys. We compare our data with theirs over the same redshift range and find them to be in good agreement (Figure \ref{agnlfcfpadovani}). 

\subsubsection{Why is our AGN RLF `high'?}
The AGN RLF for ELAIS is puzzling as it appears inconsistent with the local AGN RLF of \citet{Mauch07}. For example, the median redshift for the eight lowest luminosity AGN (the left-most three green data points in Figure \ref{agnlf}) is z~$\sim 0.05$, which is comparable to the median redshift of z~$\sim 0.07$ from \citet{Mauch07}. We specifically investigate whether these eight sources may have been mis-classified in some way. Four of the eight have clear evidence for AGN acitivity in their optical spectra (e.g. broad lines, high NII/H$\alpha$ ratios etc) and the other four have early-type spectra. The one source that also had sufficient mid-infrared data is likely to be AGN as it has mid-infrared colours consistent with other early-type galaxies. 

The AAOmega spectra are obtained through fibres 2~arcsec in diameter, which corresponds to a projected diameter of 6.4\,kpc at z=0.2. Consequently, in the cases where we detect an early-type spectrum, it is possible that we are only seeing the bulge of a star-forming galaxy \citep[see e.g.][]{Mauch07}. This is supported by the fact that the total RLF in Figure \ref{lumfn_SFvAGN} agrees very well with \citet{Mauch07} at L$_{1.4\,GHz} < 10^{22.5}$~W~Hz$^{-1}$, but the SF RLF is lower and the AGN RLF is higher. 

The next redshift slice, $0.2<\rm z \le 0.4$ is equally puzzling as it too is a factor of $\sim$ 3 higher than the local AGN RLF. Some of this may be attributed to evolution, although the large error bars preclude us from drawing any strong conclusions.  We note the further possibility that these sources are hybrid in nature, and it is possible for a source to have both a star-forming disk and a nuclear AGN. While an investigation into this would be useful, it is beyond the scope of this paper.

Another factor contributing to the higher number density of AGN may be the presence of clusters or other large-scale structures, which can have the effect of inflating the AGN RLF in ELAIS due to the small volumes probed \citep[e.g.][]{Padovani11}. For example, \citet{Mao10a} detected a large overdensity at z~$\sim 0.2$ associated with a wide-angle tailed galaxy. The removal of galaxies in this overdensity lowers the overall AGN RLF, although not by a significant amount ($\delta \Phi < 15\%$). However, there are three additional WATs in ELAIS \citep{Mao10a} at $0.3<$~z~$< 0.4$, each of which is likely to be associated with a cluster. The sampling of a larger volume would mitigate the effects of cosmic variance. 

%Finally, many calculations of the RLF have been performed on optically-selected samples \citep[e.g.][]{Mauch07}. Our data, and \citet{Padovani11} both use radio-selected samples. We do not believe the `high' AGN RLF may be attributed to selection. 

\begin{figure}
\begin{center}
\includegraphics[angle=0, scale=0.8]{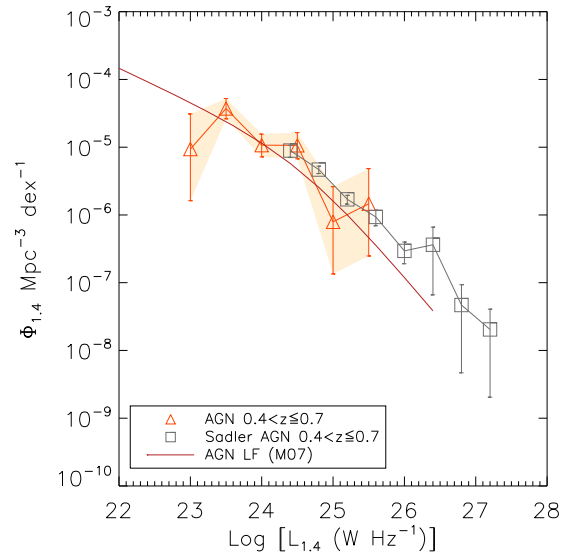}
\end{center}
\caption{Radio luminosity function for AGN in ATLAS-ELAIS with data from \citet{Sadler07} overplotted. The RLF for the redshift range $0.4<$~z~$\le 0.7$ is plotted to allow direct comparison with \citet{Sadler07}. The pink line is the local AGN RLF from \citet{Mauch07}. Errors are Poisson.}\label{agnlfcfsadler}
\end{figure}

\citet{Sadler07} calculated the RLF for AGNs at $0.4<$~z~$\le 0.7$ using radio data from FIRST and optical data from 2SLAQ. We compare our data with theirs over the same redshift range and find them to be in good agreement (Figure \ref{agnlfcfsadler}). At this redshift range we should be probing a volume large enough to sample the Universe fairly and hence be free of the effects of cosmic variance. 

We are not able to determine definitively why the AGN RLF for ELAIS appears to be `high'. We have explored the possibility of evolution, cosmic variance and the possibility of spectra being unrepresentative of the source. The AGN RLF may be suffering the effects of any one of these factors, or possibly a combination. DR3 will provide this study with more data so that we may determine why our AGN RLF is high.

\section{Summary and Conclusions}
We have presented 466 new spectroscopic redshifts for radio sources in ELAIS and CDFS as part of ATLAS, using AAOmega on the AAT. We have classified the sources as AGN or SF based on their spectra and compared these spectroscopic classifications with mid-IR diagnostics and found them to be in good agreement. Of the 466 sources, 282 are AGN, 142 are SF, 32 are either SF or AGN (none of the diagnostics used in this paper were able to determine if they were SF or AGN) and 10 are chance alignments with stars. 

We have constructed the RLF for ELAIS for both SF galaxies and AGN. We find positive evolution, consistent with previous studies, to z~=~0.5 for the SF RLF. We find our AGN RLF to be consistent with previous work by \citet{Padovani11}, and a factor of $\sim$3 higher than the work by \citet{Mauch07}. 

We cannot make any definitive statements as to why the SF RLF for ELAIS is consistent with previous studies whereas the AGN RLF for ELAIS appears inconsistent. We attribute the inconsistencies to the possibility of evolution, cosmic variance, obtaining spectra of only the bulge of a SF galaxy leading to an overestimate of early-type galaxies, or possibly an amalgamation of all these factors. We shall perform these analyses on DR3 where the extra data will allow us to draw more definitive conclusions. 
%This is most likely due to the fact that both our dataset and the dataset of \citet{Padovani11} is radio-selected whereas the dataset of \citet{Mauch07} is optically selected. A radio-selected dataset is likely to contain a higher fraction of AGN hosted by early-type galaxies. 

%tentative evidence for an overabundance of low-luminosity AGN ($<$10$^{22.5}$~W~Hz$^{-1}$) whose detection we attribute to sample variance due to the small volumes probed.% The higher number density of low-luminosity AGN potentially suggest that all AGN are radio-emitting to some degree and hence `hot-mode' accretion may be more prevalent than previously thought. 

ATLAS is the pathfinder for the forthcoming EMU Survey \citep{Norris11a}, planned for the new ASKAP telescope \citep{Johnston08}, which will survey the entire visible sky to an rms depth of 10$\,\mu$Jy\,beam$^{-1}$. Because the ATLAS and EMU surveys are well-matched in sensitivity and resolution, the results obtained in this and other papers on the ATLAS survey will be used to guide the survey design and early science papers for EMU. In particular, the spectroscopy on ATLAS sources will provide a valuable training set to guide the algorithms used to determine the redshift distribution of the anticipated 70 million EMU sources. 

\section*{Acknowledgements}
We thank the anonymous referee whose comments helped improve this paper. We thank Jamie Stevens, John Dickey, Jay Blanchard, Alastair Edge, Tom
Mauch and the ATLAS team for many fruitful discussions. We thank Paolo
Padovani for providing data for us to compare our RLFs. MYM
acknowledges the support of an Australian Postgraduate Award as well
as Postgraduate Scholarships from AAO and ATNF. NS is a recipient of
an Australian Research Council Future Fellowship. We thank the staff
at AAO and ATCA for making these observations possible. The ATCA is
part of the Australia Telescope, which is funded by the Commonwealth
of Australia for operation as a National Facility managed by
CSIRO. This research has also made use of NASA's Astrophysics Data
System.

\clearpage

\appendix

\section{Redshifts for 24\,$\mu$\MakeLowercase{m} excess sources}\label{24umsources}
Redshifts for 697 of 1080 non-ATLAS sources were obtained. 472 of these are 24$\,\mu$m  excess sources and form the basis of further investigation into the radio-FIR correlation (Norris et al., in preparation). Their redshifts were determined in the process of determining ATLAS redshifts and are presented here for the first time in Figure \ref{24umzcat}. 

\begin{table*}
\begin{center}
\caption{First 10 lines of the catalogue of new spectrosopic redshifts for 24\,$\mu$m excess sources in ATLAS. }\label{24umzcat}
\begin{tabular}{cccc}
\\
\hline

SID  & RA & Dec & z \\
& (J2000) & (J2000) & \\
\hline
SWIRE3 J033350.32-280807.2& 3:33:50.32&-28:08:07.30&      0.2885\\
SWIRE3 J033300.84-280957.3& 3:33:00.84&-28:09:57.30&      0.2147\\
SWIRE3 J033246.75-280846.9& 3:32:46.76&-28:08:47.00&      3.1880\\
SWIRE3 J033223.92-281126.8& 3:32:23.92&-28:11:26.80&      0.1961\\
SWIRE3 J033221.93-281005.8& 3:32:21.94&-28:10:05.90&      0.1811\\
SWIRE3 J033347.81-281622.0& 3:33:47.82&-28:16:22.00&      0.1029\\
SWIRE3 J033159.87-280953.0& 3:31:59.87&-28:09:53.00&      0.2363\\
SWIRE3 J033154.67-281035.9& 3:31:54.67&-28:10:36.00&      0.2148\\
SWIRE3 J033417.62-281712.8& 3:34:17.62&-28:17:12.90&      0.3400\\
SWIRE3 J033237.30-280847.2& 3:32:37.30&-28:08:47.30&      0.7683\\
\hline
\end{tabular}
\end{center}
\end{table*}

\section{V$_{enclosed}$/V$_{available}$}\label{Vmaxradio}

\begin{figure}
\begin{center}
  \includegraphics[angle=0, scale=0.8]{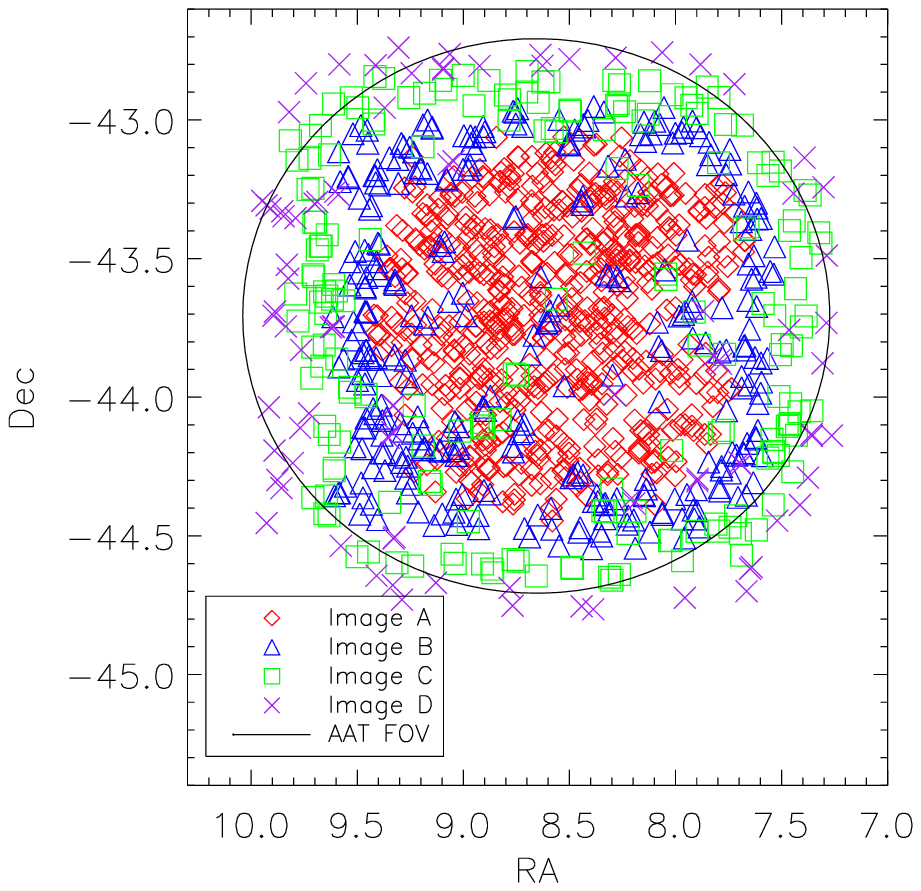}
\end{center}
\caption{RA-Dec plot of ELAIS's rms characteristics, with four different rms bins chosen, and colour-coded. To keep the plot simple, only parts of the image that aren't also part of a deeper image are plotted. That is, Image D covers the entire radio image, and Image C's datapoints can be thought of as being overlaid on Image D. The field-of-view of AAOmega is overlaid.}\label{elaisrmsAAT}
\end{figure}

The ELAIS radio image is $\sim$4.7~square degrees, of which $\pi$~square degrees were observed with the AAT. In Figure \ref{elaisrms} we presented the rms at each radio source position. Figure \ref{elaisrmsAAT} presents the same data with the AAOmega field-of-view overlaid. 

We can treat this non-uniform radio image as four overlapping radio images each with a different rms. Table \ref{elaisrmstab} presents these areas, designated images A - D with A being the deepest and D being the shallowest image. The following equations and analyses are based on the equations provided in \citet{Avni80}. 
 
\begin{table}
\begin{center}
\caption{The rms limits for the four overlapping radio images ELAIS is broken down into.}\label{elaisrmstab}
\begin{tabular}{lrr}
\\
\hline
Image & RMS limit & Area\\
&($\mu$Jy\,beam$^{-1}$)&(deg$^{2}$)\\
\hline

A & 30 & 0.95\\
B & 40 & 1.75\\
C & 65 & 2.56\\
D & 100 & 3.14\\
\hline
\end{tabular}
\end{center}
\end{table}

Although whether the source is detected is both a function of the source's position in the field, and its luminosity, the actual volume available to the source is dependent only upon its luminosity. That is to say, the source may be distributed anywhere within the total volume. This is because the volume available to the source is bound by the flux density limits presented in Table \ref{elaisrmstab}.

For a source of a given luminosity, the volume available (in Mpc$^3$) to a source in our sample is

\begin{eqnarray}
\rm V_{available}&=&\frac{1}{4\pi}\frac{4}{3}\pi(\Omega_{\rm (D-C)}\rm D_{MMAX,D}^3  \nonumber\\ &&+ \Omega_{(C-B)}\rm D_{MMAX,C}^3 \nonumber\\ & &+ \Omega_{\rm (B-A)}\rm D_{MMAX,B}^3  \nonumber\\ &&+ \Omega_{A}\rm D_{MMAX,A}^3)\\
&=&\frac{1}{3}(\Omega_{\rm (D-C)}\rm D_{MMAX,D}^3  \nonumber\\ & &+ \Omega_{\rm (C-B)}\rm D_{MMAX,C}^3 \nonumber\\ & &+ \Omega_{\rm (B-A)}\rm D_{MMAX,B}^3  \nonumber\\ & &+ \Omega_{A}\rm D_{MMAX,A}^3)
\label{Vavailableeqn}
\end{eqnarray}
where
$\Omega_{X}$ is the solid angle (in steradians) subtended by the image X, and
D$_{\rm MMAX,X}$ is the maximum comoving distance in Mpc for image X. In this way, each source is accounted for only once. This corresponds to Equation 9 of \citet{Avni80}. 

The actual volume enclosed by the source, corresponding to Equation 10 of \citet{Avni80}, is dependent on the source's redshift (and hence D$_{\rm M,source}$). For a source whose redshift is less than the maximum redshift the shallowest image (D) can detect, the enclosed volume is simply

\begin{eqnarray}
\rm V_{enclosed}&=&\frac{1}{3}(\Omega_{D} \rm D_{M,source}^3)
\label{Venclosedeqn}
\end{eqnarray}
where
D$_{\rm M,source}$ is the comoving distance in Mpc for the source. 

However, for a source whose redshift is less than or equal to the maximum redshift the deepest image (A) can detect, the enclosed volume is

\begin{eqnarray}
\rm V_{enclosed}&=&\frac{1}{3}(\Omega_{(D-C)}\rm D_{MMAX,D}^3  \nonumber\\ & &+ \Omega_{(C-B)}\rm D_{MMAX,C}^3 \nonumber\\ & &+ \Omega_{(B-A)}\rm D_{MMAX,B}^3  \nonumber\\ & &+ \Omega_{A}\rm D_{M,source}^3)
\end{eqnarray}

\begin{figure}
\begin{center}
  \includegraphics[angle=0, scale=0.8]{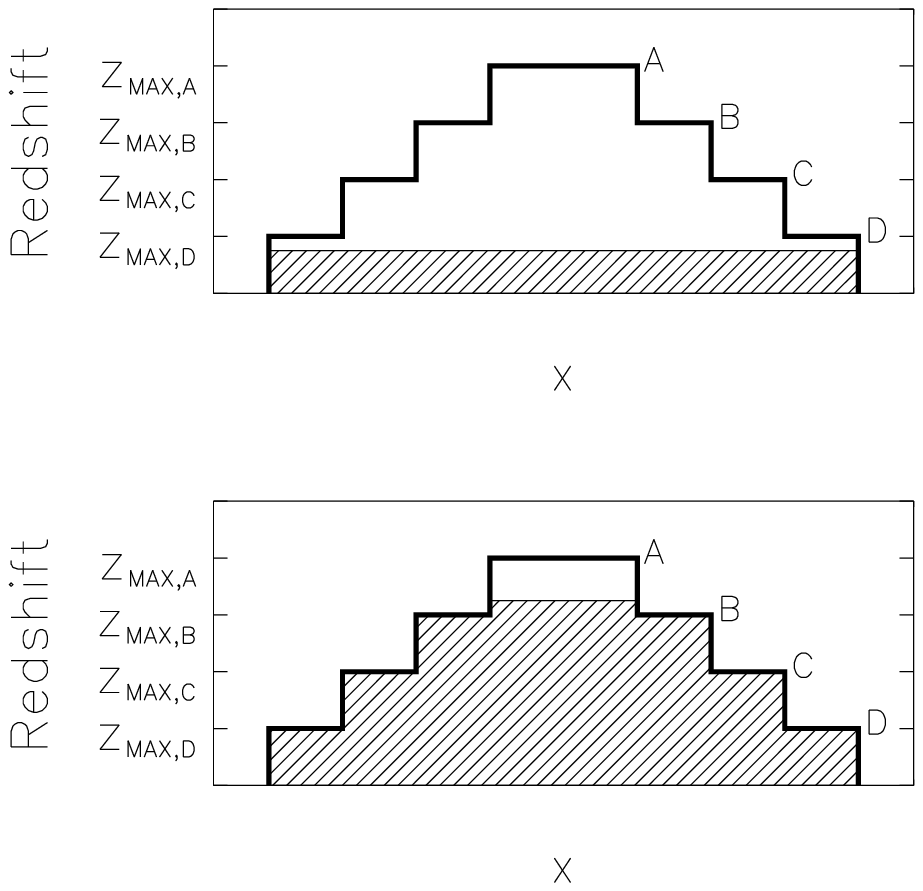}
\end{center}
\caption{A simple model depicting V$_{available}$ (solid black line) and V$_{enclosed}$ (shaded region) for a source whose redshift is less than the maximum redshift detectable in the shallowest image (upper panel) and for a source whose redshift is less than the maximum redshift detectable in the deepest image (lower panel). The x-axis is an idealised cross-section of the area and the y-axis is the redshift. }\label{schematic}
\end{figure}

This is shown schematically in Figure \ref{schematic}. These are for the two extremes but the same procedure is followed for sources with redshifts between z$_{MAX,C}$ and z$_{MAX,D}$ as well as for sources with redshifts between z$_{MAX,B}$ and z$_{MAX,C}$.

%\subsection{SF v AGN using the FRC and maybe polarisation?}
%\begin{figure}
%\begin{center}
%\includegraphics[angle=0, scale=0.45]{../..//atlas_spectroscopy/luminosity/flux24_S20.ps}
%\end{center}
%\caption{ }\label{frc}
%\end{figure}

%Do a quick and dirty SF and AGN split using the FRC. ie. If it's off the correlation it's an AGN, otherwise it's SF. 

%Propose to do this with multiple diagnostics in Paper 2?

\label{lastpage}
\end{document}